\documentclass{emulateapj}

\newcommand\degree{{^\circ}}

\newcommand\kpc{{\rm\,kpc}}

\newcommand\kmsec{{\rm\,km\,s^{-1}}}
\newcommand\kms{\kmsec}

\newcommand\surfb{{\rm\,mag\,arcsec^{-2}}}

\shorttitle{The Kinematics of Thick Disks}
\shortauthors{Yoachim \& Dalcanton}

\begin{document}

\slugcomment{ {\it Accepted for publication in the Astrophysical Journal}}
\title{The Kinematics of Thick Disks in External Galaxies}

\author{Peter Yoachim\altaffilmark{1} 
  \&  Julianne J. Dalcanton\altaffilmark{2,3}}
\affil{Department of Astronomy, University of Washington, Box 351580,
Seattle WA, 98195}

\altaffiltext{1}{e-mail address: {yoachim@astro.washington.edu}}
\altaffiltext{2}{Alfred P. Sloan Foundation Fellow}
\altaffiltext{3}{e-mail address: {jd@astro.washington.edu}}

\begin{abstract}
We present kinematic measurements of the thick and thin disks in two
edge-on galaxies.  We have derived stellar rotation curves at and
above the galaxies' midplanes using Ca {\sc ii} triplet features
measured with the GMOS spectrograph on Gemini North.  In one galaxy,
FGC~1415, the kinematics above the plane show clear rotation that lags
that of the midplane by $\sim$20-50\%, similar to the behavior seen in
the Milky Way.  However, the kinematics of the second galaxy, FGC~227,
are quite different.  The rotation above the plane is extremely slow,
showing $\lesssim$25\% of the rotation speed of the stars at the
midplane.  We decompose the observed rotation curves into a
superposition of thick and thin disk kinematics, using 2-dimensional
fits to the galaxy images to determine the fraction of thick disk
stars at each position.  We find that the thick disk of FGC~1415
rotates at 30-40\% of the rotation speed of the thin disk.  In
contrast, the thick disk of FGC~227 is very likely counter-rotating,
if it is rotating at all.  These observations are consistent with the
velocity dispersion profiles we measure for each galaxy.  The
detection of counter-rotating thick disks conclusively rules out
models where the thick disk forms either during monolithic collapse or
from vertical heating of a previous thin disk.  Instead, the data
strongly support models where the thick disk forms from direct
accretion of stars from infalling satellites.
\end{abstract}


\section{Introduction}
Since the discovery of the Milky Way's thick disk in 1983
\citep{Gilmore83}, there has been a growing body of evidence that the
thick disk is a structurally, chemically, and kinematically distinct
component of the Galaxy.  Structurally, the Milky Way's thick disk has
a significantly larger scale height than the thin disk (for reviews of
the thick disk scale height see \citet{Reid93, Buser99, Norris99} and
references therein), and a possibly somewhat longer scale length
\citep{Robin96, Ojha01, Chen01, Larson03}.  Chemically, thick disk
stars are more metal-poor and older than stars in the thin disk
\citep{Reid93, Chiba00}.  They are also significantly enhanced in
$\alpha$-elements, compared to thin disk stars of comparable iron
abundance \citep{Prochaska00, Taut01, Bensby03, Feltzing03,
Mishenina04, Brewer04}.  Kinematically, thick disk stars have both a
larger velocity dispersion and slower net rotation than stars in the
thin disk \citep{Nissen95, Chiba00, Gilmore02, Soubiran03, Parker04}.  All of
these facts lead to the conclusion that the thick disk is a relic of
the young Galactic disk.  As such, it provides an excellent probe of
models of disk galaxy formation (see recent reviews by
\citet{Nissen03, Freeman02}).

Originally detected in S0 galaxies \citep{Burstein79, Tsikoudi79},
thick disk components have since been detected in many galaxies,
including S0's \citep{deGrijs96, deGrijs97, Pohlen04}, Sb's
\citep{Kruit84, Shaw89, vanDokkum94, Morrison97, Wu02}, and later type
galaxies \citep{Abe99, Neeser02, Yoachim04}.  Across all Hubble types,
photometric decompositions consistently find that thick disk scale
heights are 2-6 times larger than thin disk scale heights and that the
thick disks' scale lengths are slightly larger (\citet{Shaw89, Wu02,
Neeser02, Pohlen04}, but see also \citet{Abe99}).

Thick and thin disks are likely to have distinct formation mechanisms,
given the systematic differences between their properties.  The
structure, dynamics, and chemical abundance of the thin disk strongly
suggest that the majority of its stars formed gradually from a
rotating disk of high angular momentum gas \citep{Fall80, Matteucci89,
Chiappini97}.  In contrast, the formation of the thick disk is still
poorly understood and is likely to be more complex.  The large scale
height of the thick disk suggests that its stars were either (1)
vertically ``heated'' from a previously thinner disk \citep{Quinn93,
Velazquez99, Robin96, Chen01}, (2) formed from gas with a large scale
height \citep{ELS, Norris91, Burkert92, Kroupa02, Fuhrmann04,
Gilmore86}, or (3) directly deposited at large scale heights during
the accretion of smaller satellite galaxies \citep{Bekki01, Gilmore02,
Abadi203, Martin04, Navarro04, Brook04}.  Most models of thick disk
formation fall into one of these three cases.  In the first, the thick
disk stars form initially in a thin disk.  In the second, they form
within the thick disk itself.  In the third, they form outside of the
Galaxy entirely.

All three of the above scenarios are compatible with the structure of
the Milky Way's thick disk.  However, they all imply very different
behavior for the early evolution of disk galaxies.  The first case
would suggest that disks form largely through smooth gas accretion
that is occasionally punctuated by minor merging events.  The second
case would suggest either that disks form primarily through smooth
monolithic collapse, with thick disk stars precipitating out of the
collapsing gas cloud \citep{ELS}, or that the thick disk forms from a
disk of gas that has been energetically heated by star formation
\citep{Kroupa02}.  The third case would suggest that disks can form
entirely from merging sub-units, in spite of their highly ordered
present day structure.  The degeneracies between these models must be
broken before the properties of the thick disk can be used as a
constraint on theories of galaxy formation.

Past attempts to distinguish among formation models have often relied
on the chemical abundances of thick disk stars.  The low metallicities
of thick disk stars suggest that they formed early in the evolution of
the Galaxy.  Their $\alpha$-enhancement suggests that the time scale
for their formation was sufficiently rapid to suppress iron enrichment
by Type Ia supernovae.  Unfortunately, the resulting implication that
the thick disk formed early and rapidly can be accommodated by all of
the possible formation scenarios.  \citet{BlandHawthorn04} suggest
that metallicity measurements of more than $10^5$ stars would be
required to definitively constrain the origin of the Milky Way's thick
disk.  Metallicities of thick disk stars therefore remain a somewhat
weak discriminant among competing models.

In contrast to metallicities, stellar kinematics show more promise for
constraining the origin of the thick disk.  If the thick disk results
from dynamical heating of a young previously thin disk, both
components should have similar rotation curves, assuming similar
angular momenta for the gas that falls in before and after the
heating event.  If the thick disk forms from gas as it collapses into
the galaxy, then angular momentum conservation requires that the
rotation of the thick disk must always lag the rotation of the thin
disk, by an amount that depends on the relative scale lengths of the
thick and thin disks.  Finally, if the thick disk is assembled from
stars originally formed in merging satellites, then the relative
kinematics of the thick and thin disks are likely to be highly
variable.

The rotation of the Milky Way thick disk is marginally consistent with
all of these scenarios \citep{Burkert92, Robin96, Buser99, Gilmore02,
Soubiran03}. Measurements have consistently shown the Galactic thick
disk to be co-rotating with the thin disk, but with modest lag.  The
exact value of the lag varies by a factor of two depending on the
study, ranging from $\sim40$ km s$^{-1}$ \citep{Reid98, Chiba00} up to
$\sim90$ km s$^{-1}$ \citep{Gilmore02, Parker04}.  The smaller
rotational lags can be easily produced in most thick disk formation
models, but the larger lags observed recently are more difficult to
explain without satellite accretion.  However, rather than ruling out
out alternative models, the larger lags are often interpreted as
contamination from a single disrupted satellite.

Although data within the Galaxy have not been sufficient to
definitively constrain the origin of the thick disk, the Milky Way is
only one single case.  Instead, measurements of thick disk kinematics
in {\emph{several}} galaxies are necessary to exploit the full power
of kinematic discriminants among formation models.  Scenarios
involving mergers will naturally lead to much more kinematic variation
among the population of thick disks than if thick disks formed during
monolithic collapse.  Even among merging models, we would expect the
relative kinematics of the thick and thin disks to vary more if the
thick disk stars are directly accreted than if the thick disk was
built from vertically heated thin disk stars.  By measuring the
relative kinematics of thick and thin disks in a large sample of
galaxies, one can therefore discriminate among thick disk formation
models far more effectively than by studying the Milky Way alone.

In this paper we present the first results of our on-going study of
the kinematic properties of thick disks in a large sample of late-type
edge-on galaxies \citep{Dalcanton00, Dalcanton02}.  We have measured
of the kinematics of thick and thin disk stars in two edge-on
late-type disk galaxies.  We first present the sample selection and
observations in \S\ref{observationsec}.  We then describe the data
reduction and extraction of the rotation curves in
\S\ref{reductionsec}.  In \S\ref{RCsec} we decouple the observed
rotation curves to derive the rotation of the thick and thin disks.
In \S\ref{losvdsec} we measure radially resolved velocity dispersion
profiles above and below the plane.  We then discuss the implication
of our initial results for models of disk galaxy formation in
\S\ref{discussionsec} and summarize our conclusions in
\S\ref{conclusionsec}.

\section{Observations}                \label{observationsec}

We have carried out long-slit spectroscopic observations of two
galaxies drawn from the \citet{Dalcanton00} sample of edge-on
late-type galaxies.  The original sample of 49 galaxies was selected
from the Flat Galaxy Catalog \citep{Karachentsev93} and imaged in $B$,
$R$, and $K_s$ \citep{Dalcanton00}.  \citet{Dalcanton02} used this
imaging to demonstrate the ubiquity of thick disks around late-type
galaxies.  We have since used two-dimensional decompositions of the
galaxy images to measure structural parameters for the thick and thin
disks.  This decomposition will be presented in detail in a future
paper \citep{Yoachim04}, and we now give a brief overview of our
decompositions.

\begin{figure}
\plotone{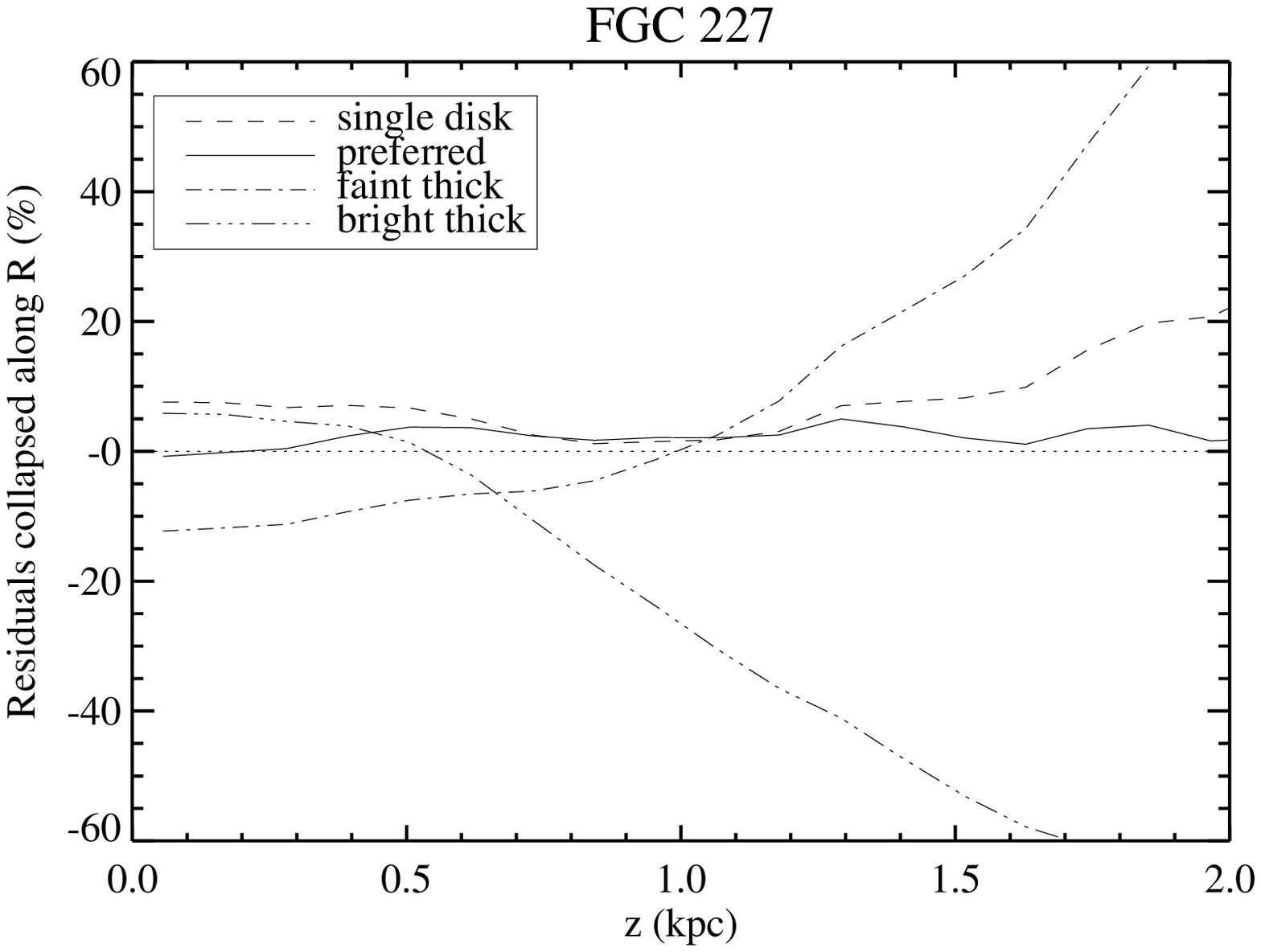}
\plotone{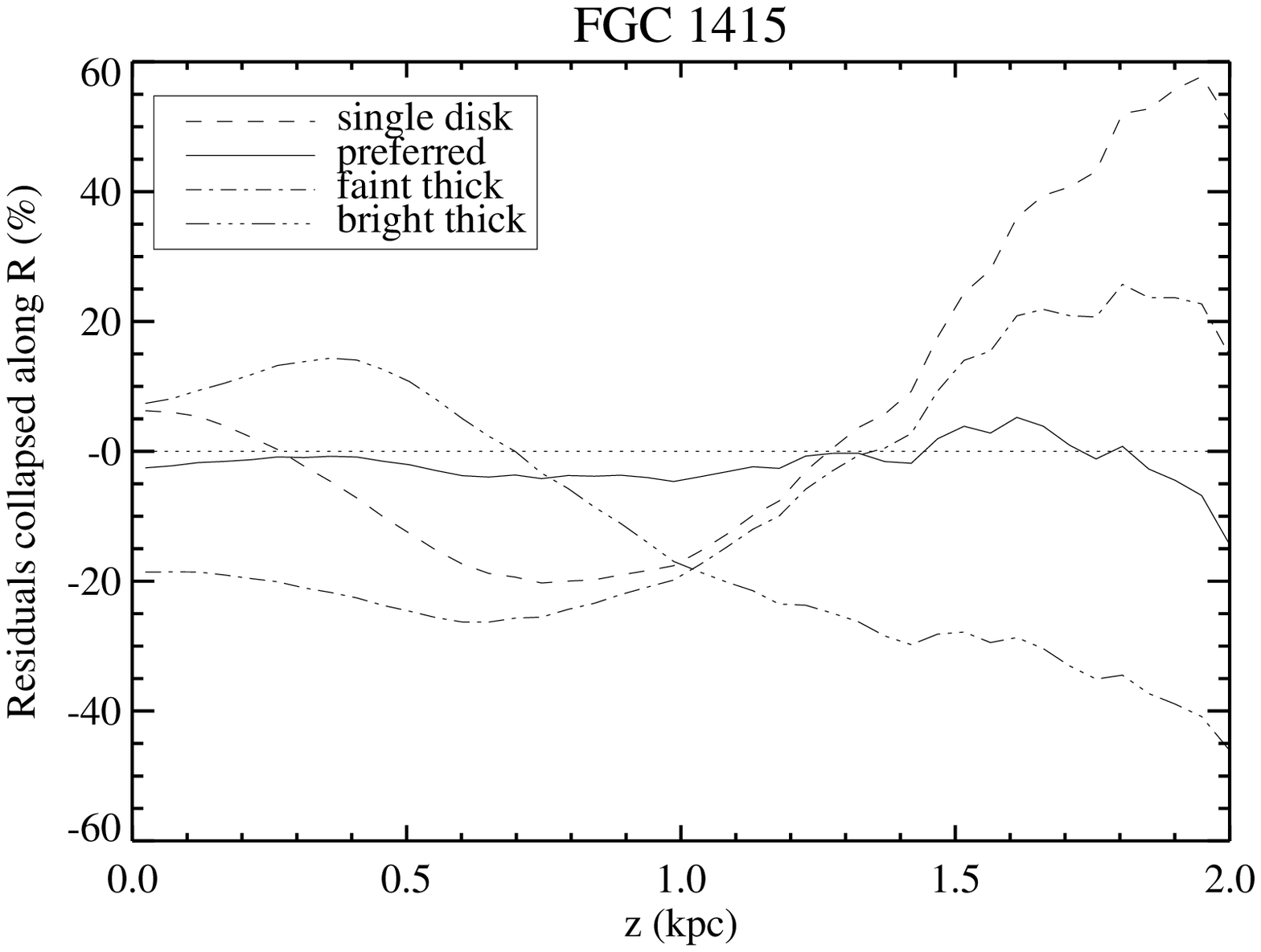} 
\caption{Vertical residual profiles for photometry fits to our
galaxies.  Plots were made by collapsing residual images along the
major axis and normalizing by the model galaxy and averaging both
sides of the galaxy. In both galaxies, the single disk provides a poor
fit and leaves excess flux at high galactic latitudes.  Changing the
functional form of the single disk does not alleviate this problem.
We have also plotted the residuals from our extreme models.  As
expected, the ``faint thick'' model leaves excess flux at large $z$,
while the ``bright thick'' over-subtracts at high $z$.
\label{resids} }
\end{figure}

When fitting models to our galaxies, we adopt the method of
\citet{Kregel02} and use a Levenberg-Marquardt $\chi^2$ minimization
routine.  Each disk is modeled as having a surface brightness 
\begin{equation}
\Sigma(R, z) = \Sigma_{0}
(R/h_{R})K_{1}(R/h_{R}) \mathrm{sech}^{2}(z/(2h_z))
\end{equation}
where $K_{1}$ is a modified Bessel function of the first order, $
\Sigma_{0}$ is the edge-on central surface brightness, $R$ is the
projected radius along the major axis, and $h_z$ is the exponential
scale height.  We weight individual pixels by the inverse
\emph{model}.  This method allows one to fit regions of very low
surface brightness without being overly skewed by brighter midplane
features.  Comparing the formal reduced $\chi^2$ values of the single
disk and two-disk models using an $F$-test \citep{Matthews00}, we find
that the two-disk models are strongly justified.  To derive estimates
for our parameter uncertainties, we fit a series of models varying the
functional form of the vertical profiles (e.g. sech($z$), sech$^2(z)$)
for both the thick and thin disk as well as comparing models which
were fitted with the midplane region masked.  We estimate our
uncertainties by using the full range of convergent values in all the
models.  Our final fitted parameters and uncertainties are listed in
Table~\ref{galtable}.  Plots of our residuals collapsed along the
major axis are shown in Figure~\ref{resids}.  Our residuals clearly
show that subtracting off a single disk leaves excess flux at high $z$
while our preferred models fit well across the full range of
latitudes.  This procedure is discussed more extensively
in~\citet{Yoachim04}.

\begin{deluxetable*}{lcccccccccc}
\tabletypesize{\scriptsize}
\tablecaption{Properties of Targeted Galaxies \label{galtable} \\
{\scriptsize Columns: (1) Galaxy name from the Flat Galaxy Catalog;
  (2) Measured Heliocentric recessional velocity; (3) Adopted distance
  from \citet{Kara00}; (4) 21-cm line width at 50 percent of the peak;
  (5) Thin disk edge-on $R$-band central surface brightness; (6) Thin
  disk exponential scale length (Structural parameters taken from
  \citet{Yoachim04}); (7) Thin disk exponential scale height; (8)
  Thick disk edge-on central surface brightness; (9) Thick disk scale
  length; (10) Thick disk scale height; (11) Ratio of the thick disk
  luminosity to the total galaxy luminosity.}
 }
\tablewidth{0pt}
\tablehead{
\colhead{Galaxy} & 
\colhead{$V_r$} & \colhead{$H_0 d$} & \colhead{W$_{50}$/2} & 
\colhead{$\mu_{0,thin}$} & \colhead{$h_{r,thin}$} & \colhead{$h_{z,thin}$} & 
\colhead{$\mu_{0,thick}$} & \colhead{$h_{r,thick}$} & \colhead{$h_{z,thick}$} &
\colhead{$L_{thick}/L_{total}$} 
\\
\colhead{} & 
\colhead{$\kms$} & \colhead{$\kms$} & \colhead{$\kms$} & 
\colhead{mag/$\sq\arcsec$} & \colhead{$(\arcsec)$} & \colhead{$(\arcsec)$} & 
\colhead{mag/$\sq\arcsec$} & \colhead{$(\arcsec)$} & \colhead{$(\arcsec)$} & 
\colhead{}
\\
\colhead{(1)} & 
\colhead{(2)} & \colhead{(3)}  & 
\colhead{(4)} & \colhead{(5)} & 
\colhead{(6)} & \colhead{(7)} & 
\colhead{(8)} & \colhead{(9)} &
\colhead{(10)} & \colhead{(11)}
}
\startdata
FGC 227  & 5493 & 6261 &    107  &21.5$\pm0.5$ & 10.0$\pm0.8$ & 1.0$\pm0.2$ & 22.8$\pm0.8$& \phn9.3$\pm1.3$ & 2.0$\pm0.4$ & 0.50 \\
FGC 1415 & 1522 & 2683 & \phn87  & 20.9$\pm0.3$ & 14.8$\pm0.9$ & 1.2$\pm0.3$ &  22.2$\pm0.3$&   20.0$\pm1.2$ & 3.2$\pm0.3$ & 0.44 \\
\enddata
\end{deluxetable*} 

\begin{figure*}
\epsscale{.7}
\plotone{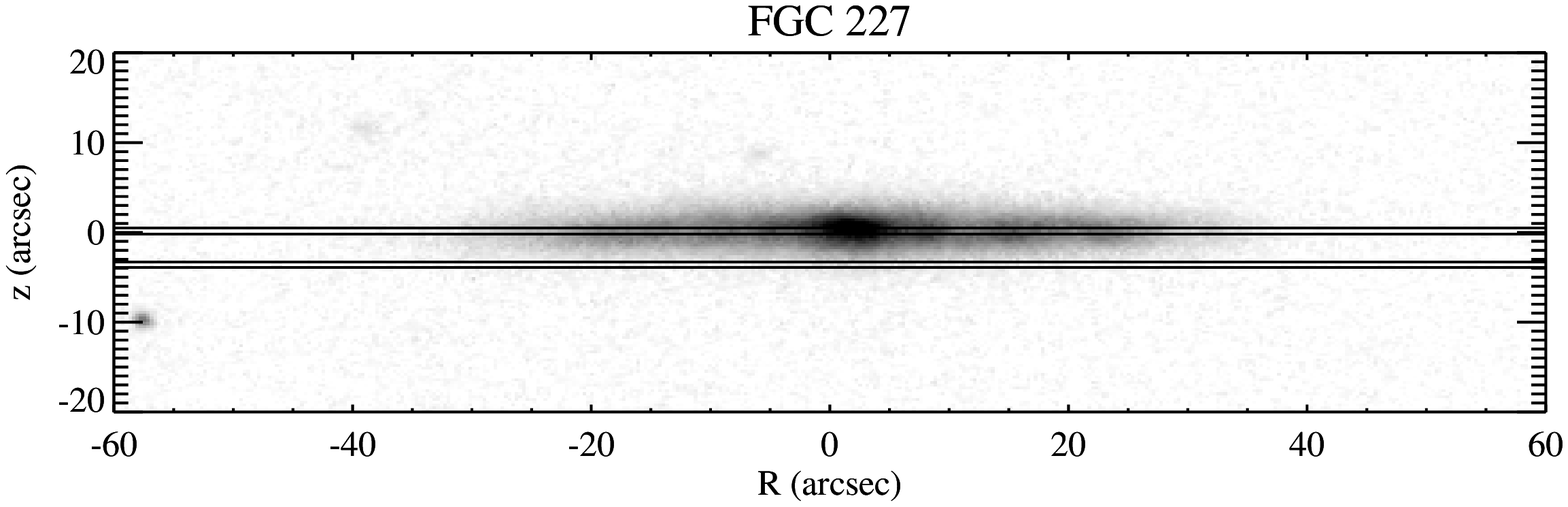} 
\plotone{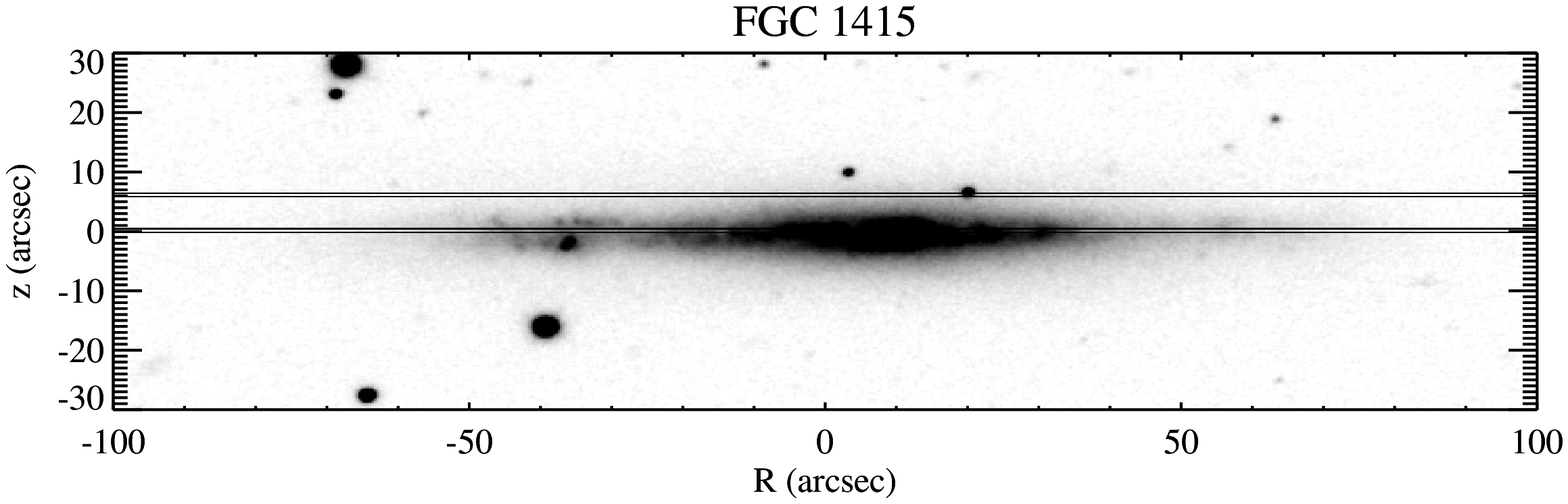} 
\caption{$R$-band images of our galaxies. Solid lines represent the
  location of the slit jaws for our midplane and off-plane
  observations.\label{slit_pos} }
\end{figure*}

To define a sample for measuring the thick disk kinematics, we
selected several initial candidates for spectroscopic follow-up.  We
chose to focus our initial efforts on galaxies with rotation speeds
near $V_c\sim100\kms$.  These galaxies are rotating fast enough that
their kinematics can be measured with moderate resolution
spectroscopy.  However, they are also low enough mass that they fall
below the $V_c\sim120\kms$ transition where dust lanes become common
\citep{Dalcanton04}, allowing us to probe their thin disk kinematics
and structure more reliably.  From the galaxies with $V_c\sim100\kms$,
we further restricted our sample to the most nearby, spatially
well-resolved galaxies.  The resulting sample was then submitted to
the NOAO Gemini North observing queue.  The choice of the specific
galaxies targeted was then left to the Gemini observers, to maximize
the chances that one of our targets would be visible at any night
during the observing semester.  In this paper we present results for
the two galaxies that were selected by the Gemini queue -- FGC~227
and FGC~1415.  The properties of both galaxies are listed in Table 1,
details of our observations are listed in Table~\ref{obstable}, and
images of the galaxies are shown in Figure~\ref{slit_pos}.

\begin{deluxetable*}{lccccccc}
\tabletypesize{\scriptsize}
\tablecaption{Observing Details\label{obstable}}
\tablewidth{0pt}
\tablehead{
\colhead{Galaxy} & 
\colhead{RA} & \colhead{Dec} & 
\colhead{Slit PA} & \multicolumn{2}{c}{Offset from midplane} & 
\colhead{\% of Light from} &
\colhead{Integration}
\\
\colhead{} & 
\colhead{(J2000)} & \colhead{(J2000)} & 
\colhead{$(\deg$ E of N)} & \colhead{$(\arcsec)$} & \colhead{$(\kpc)$} & 
\colhead{Thick Disk} &
\colhead{(minutes)} 
}
\startdata
FGC 227  & 02:00:56.7 & 19:42:26 & 9 & -   & -   & 20\% & \phn3x20 \\
FGC 227  &  02:00:56.5 & 19:42:25 & 9 & 3.0 & 1.3 & 50\% &    28x20 \\
FGC 1415 & 12:20:27.7 & 01:28:09 & 146 & -   & -   & 30\%      & \phn3x20 \\
FGC 1415 & 12:20:28.0 & 01:28:12 & 146 & 5.4 & 1.0 & 70\%      &    46x20 \\
\enddata

\end{deluxetable*}

We obtained long-slit spectroscopy of the Ca {\sc ii} triplet (8498\AA,
8542\AA, and 8662\AA) to measure stellar kinematics.  We chose
two slit locations for each galaxy, oriented with the slit parallel the
major axis of the galaxy.  The first slit position was located along
the midplane, where the light is dominated by thin disk stars.  The
second slit position was located 3--4.5 thin disk scale heights above
the midplane, where $50-80$\% of the light came from the thick disk,
based on the thick and thin disk decompositions of \citet{Yoachim04}.
The locations of the slits are superimposed on the galaxy images shown
in Figure~\ref{slit_pos}.  Typical surface brightnesses in the
midplane and high-latitude slit positions were 21.5$\surfb$ and
23$\surfb$, respectively, in the $R$-band.

The galaxies were observed with the Gemini North GMOS spectrograph
using an 0.5$\arcsec$ wide slit and the R400-G5305 grating set
with a central wavelength of $\sim\!8440$\AA.  The CCD was binned by 2
in the spatial direction during readout.  The resulting spectra have
spectral resolution of 3.4\AA\ (FWHM) near the Ca {\sc ii} triplet region as
measured by CuAr arc spectra.  This resolution allows lines to be
centroided to $\sim$10$\kms$.  

Accurate placement of our slit is of great importance because we are
combining images taken over several nights.  Our offsets from the
midplane are relatively small, and an error in placement of order an
arcsecond could compromise our results.  In our remote observing
program, we instructed the Gemini observing specialists to align the
slit along the galaxy midplane, after which, a pre-programmed offset
iterator would shift the slit to the appropriate off-plane locations.
To place the slit, the observing specialist first took an exposure
with no dispersion element to locate the galaxy midplane.  The
longslit was then moved into the light path and another exposure was
taken.  If the slit did not fall on the galaxy midplane, the telescope
was offset to the correct position.  Inspecting the returned
acquisition images, we find that the observing specialists were able
to accurately place the slit along the midplane within $\sim0.2$
arcseconds (1-2 GMOS pixels).  Because our spectra are dithered along
the slit with offsets of around 5-10 arcseconds, we have checked the
overall accuracy of Gemini offset commands and found that they are
correct to $\pm1$ pixel (0.15 arcseconds).  Overall, we find that the
slit was always placed within 0.21$\arcsec$ of our target position which
ensures our target regions were always in the 0.5$\arcsec$ slit.

Observations of FGC~227 and FGC~1415 were executed during the spring
and fall observing semesters of 2003, respectively.  The image quality
was in the 70th-85th percentile, corresponding to typical seeing of
$\sim1\arcsec$.  Because our galaxies are large ($\sim1\arcmin$), as
are the offsets between the two slit observing positions
($\gtrsim3-5.5\arcsec$), poor seeing has little effect on the
resulting spectra.  The midplane slit positions were observed with
three 20~minute exposures for both galaxies.  The thick disk of
FGC~1415 was observed in 46$\times$20~minute exposures over 9 separate
nights, for a total of 15.3 hours.  The thick disk of FGC~227 was
observed in 28$\times$20~minute exposures over 8 separate nights, for
a total of 9.3 hours.  The spectra were dithered along the slit in a
random pattern between exposures, with typical shifts of 5$\arcsec$.

\section{Data Reduction}            \label{reductionsec}

The GMOS CCD's were bias corrected using both bias images and fits to
31 columns of overscan.  Flat fielding was performed with the Facility
Calibration Unit (FCU) and the twilight sky.  FCU flats were
interleaved with the science exposures approximately every hour to
ensure complete removal of any variation in the CCD fringing pattern.
Twilight flats were combined and used to correct for uneven
illumination of the CCD.

The spectra were wavelength calibrated using night sky emission lines.
We combined the sky line data of \citet{Oster96} with the IRAF sky
line list to construct an atlas of lines (or line blends) that would
be bright and centroidable at our instrumental resolution.  This
procedure yielded $\sim\!40$ lines for chip 2 (containing the Ca {\sc ii}
triplet) and $\sim\!30$ lines for chip 3.  We fit
the resulting wavelength solution with a 6th degree Chebychev
polynomial.  RMS errors in wavelength calibration were of order
0.3 \AA, or $\sim\!10\kms$ at the wavelengths of interest.

Analysis of the Ca {\sc ii} triplet absorption line features requires accurate
sky subtraction due to the high density of sky lines at long
wavelengths.  Each column of the wavelength-rectified spectrum was fit
with a low order ($n=1-4$) polynomial in regions uncontaminated by the
galaxy.  The resulting polynomial was then subtracted from the entire
column.  Unfortunately, in our deepest off-plane integrations, clear
errors in sky-subtraction remain.  These errors are quite small
(typically only 5\% of the sky), and are not noticible in low
signal-to-noise images.  However, they are a substantial source of
uncertainty in the off-plane spectra, where the surface brightness of
the galaxy is less than 10\% of the sky and only 1\% of the brightest
skylines.  Measurements of an image of the slit suggest that these
errors result from small-scale (10\%) variations in slit width, that
lead to systematic errors when removing extremely bright sky lines.
While nod-and-shuffle techniques can greatly reduce this problem,
our targets are too large to effectively use this observing mode.

After sky subtraction, frames were corrected for atmospheric
extinction.  No absolute flux calibration was applied.  The spectra
were then Doppler corrected to the local standard of rest, spatially
aligned, and combined using a sigma-clipping algorithm to eliminate
cosmic rays.

Before extracting one-dimensional spectra, we re-binned our galaxy
spectra to a scale linear in $x= \ln \lambda$. Such re-binning allows
Doppler velocity shifts to be performed as linear shifts in $x$.  The
logarithmic bin size was set to 25 km/s in order to preserve the total
number of pixels.

\begin{figure*}
\epsscale{.6}
\plotone{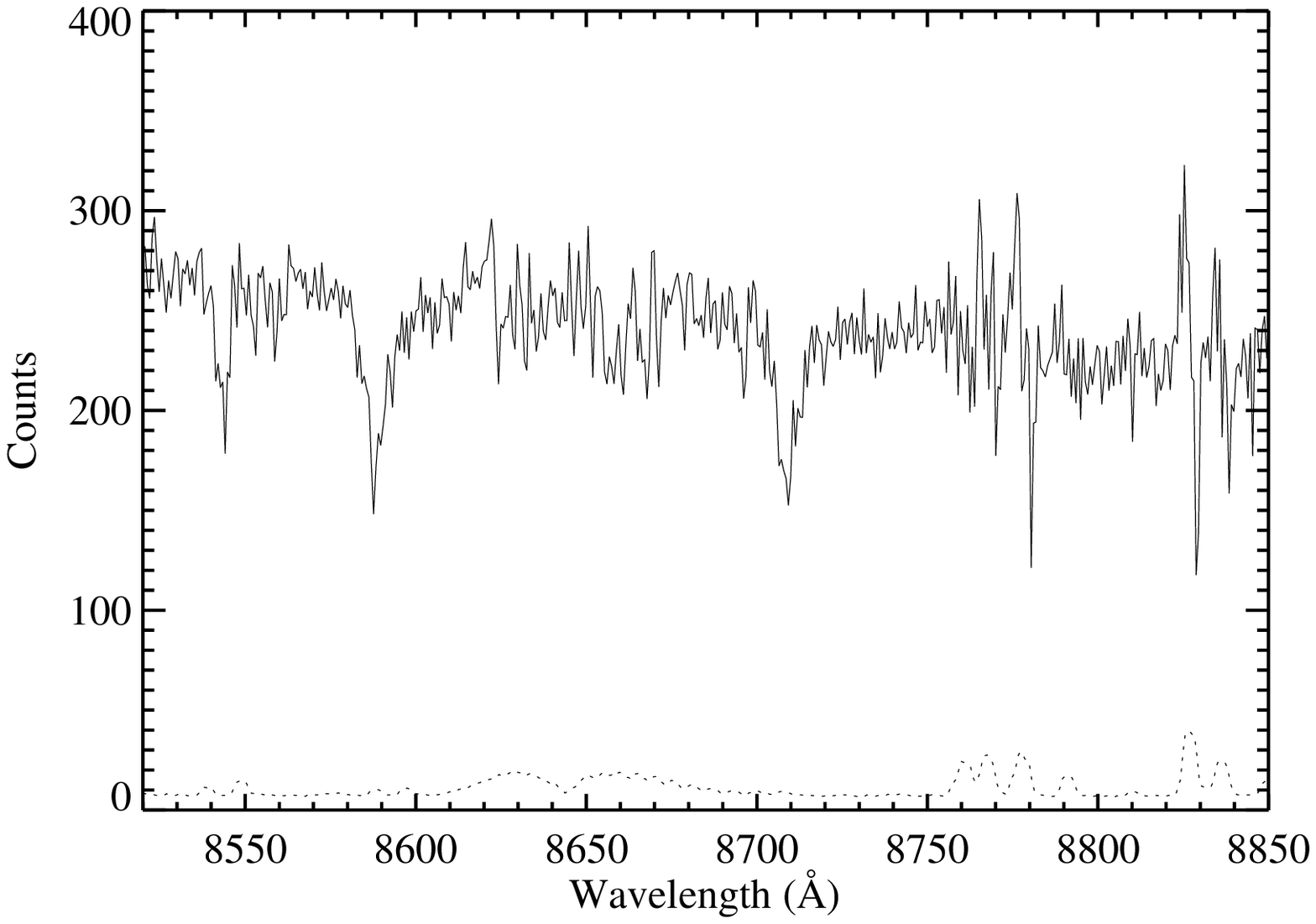}\\ 
\plottwo{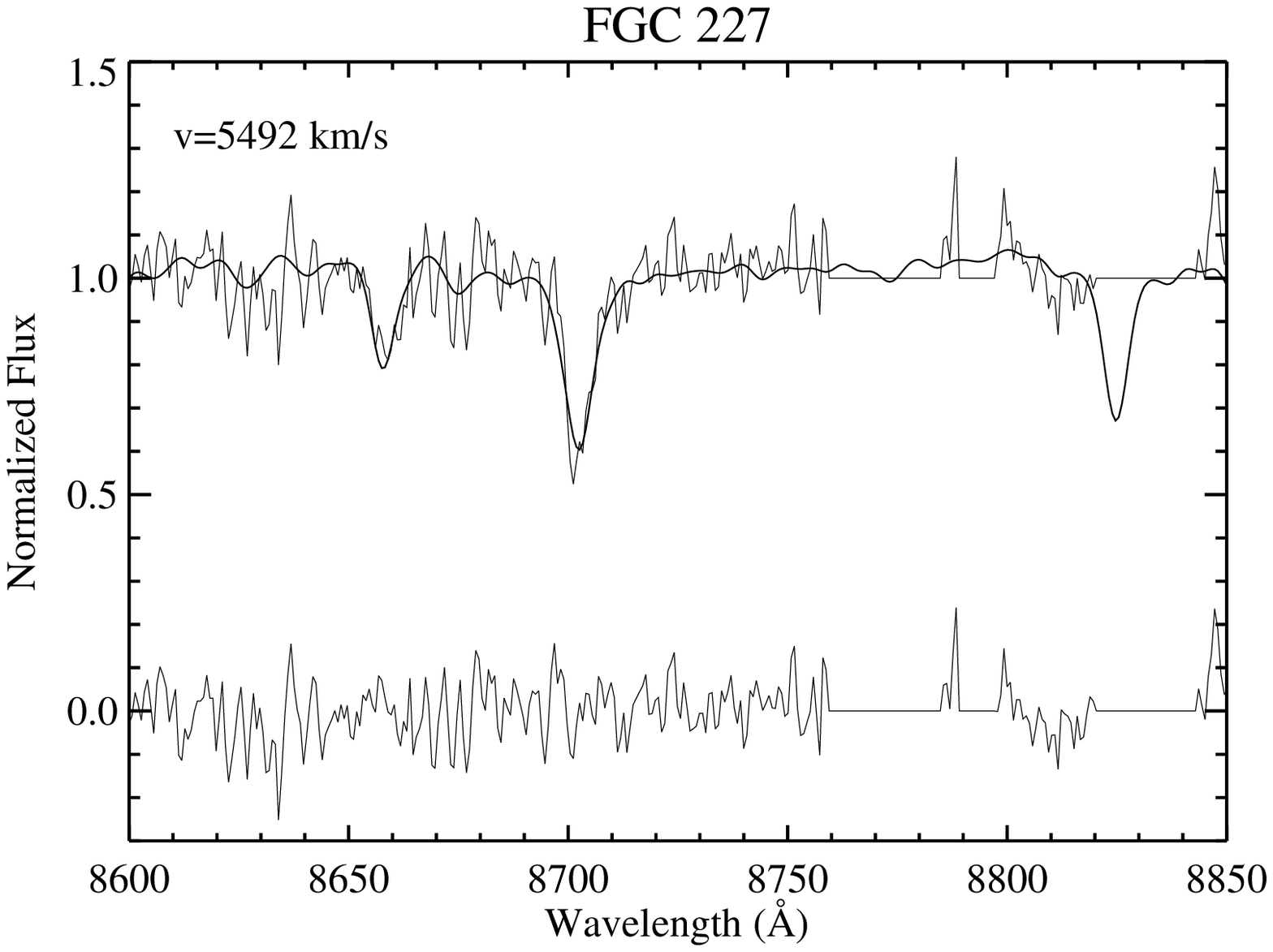}{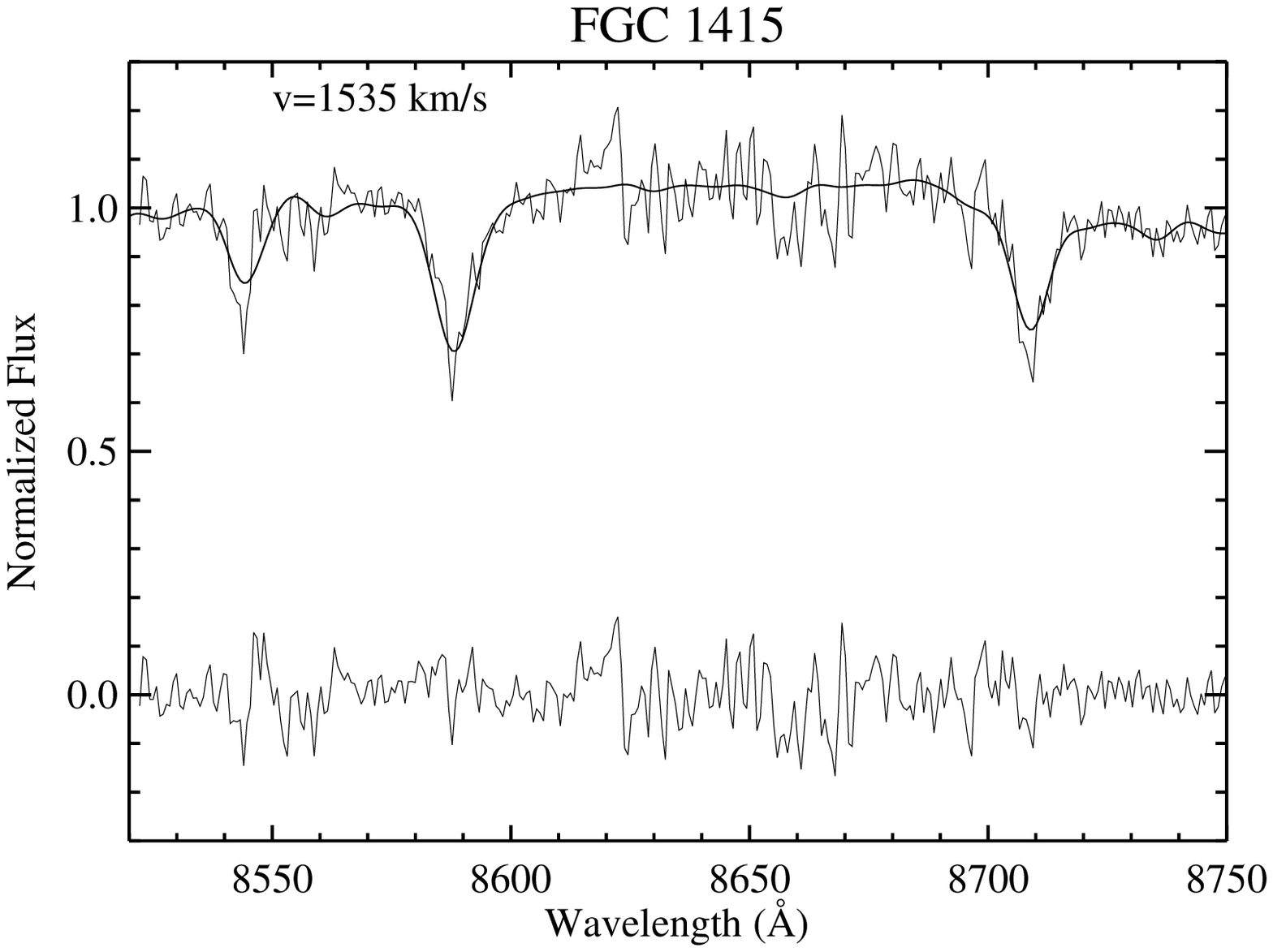} 
\caption{(Top) Raw extracted galaxy spectra (solid) along with the
Poisson noise from the sky spectrum (dotted).  (Bottom) Example
extracted spectra of both midplanes, skyline residuals have been
masked for FGC 227 resulting on one of the Ca triplet lines being
lost.\label{ex_fit} }
\end{figure*}

Spectra were extracted by summing 30-300 pixels perpendicular to the
dispersion direction, corresponding to radial bins of 4-40$\arcsec$.
We varied the extraction bin size to ensure that each spectra reaches
an average signal-to-noise ratio of 10-15 per pixel.  In the case of
FGC~227's shallower off-plane observations, we were forced to bin to a
lower S/N ($\sim\!5$) to obtain reasonable sampling across the length
of the galaxy.  Once extracted, the galaxy spectra were normalized by
dividing by a low-order polynomial.  Error spectra were computed by
analyzing regions of blank sky on the reduced images, and are clearly
dominated by the sky-subtraction problems discussed above (see
Figure~\ref{ex_fit}).  In regions where the sky spectra is smooth, our
noise spectra closely matches expectations from Poisson statistics and
read noise.  Columns that showed excessive systematic residuals caused
by skylines ($> 5 \sigma$ above what would be expected from Poisson
noise) were flagged and masked from later analysis.

\section{Template Fitting}             \label{RCsec}

\begin{figure*}
\epsscale{1}
\plottwo{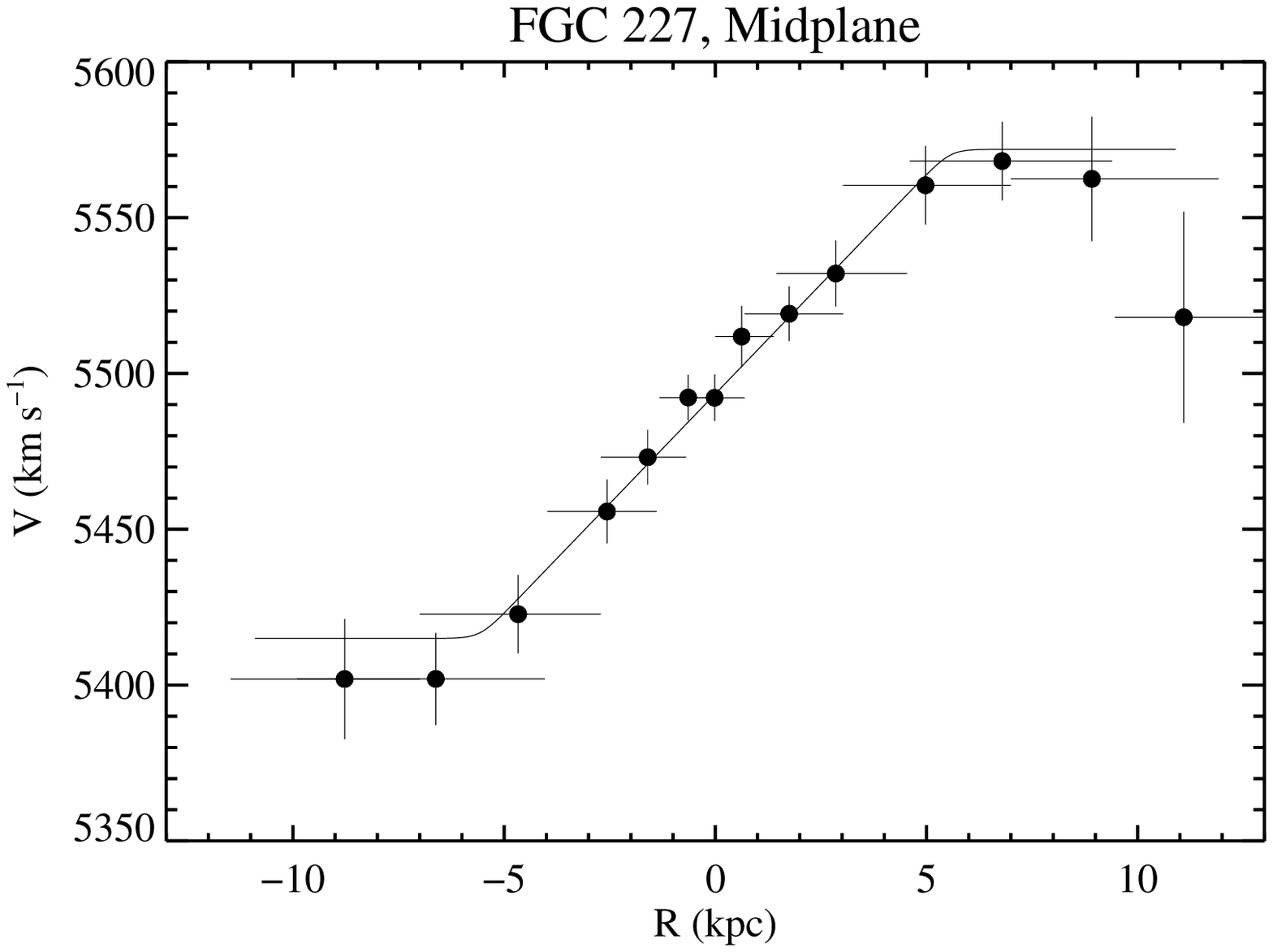}{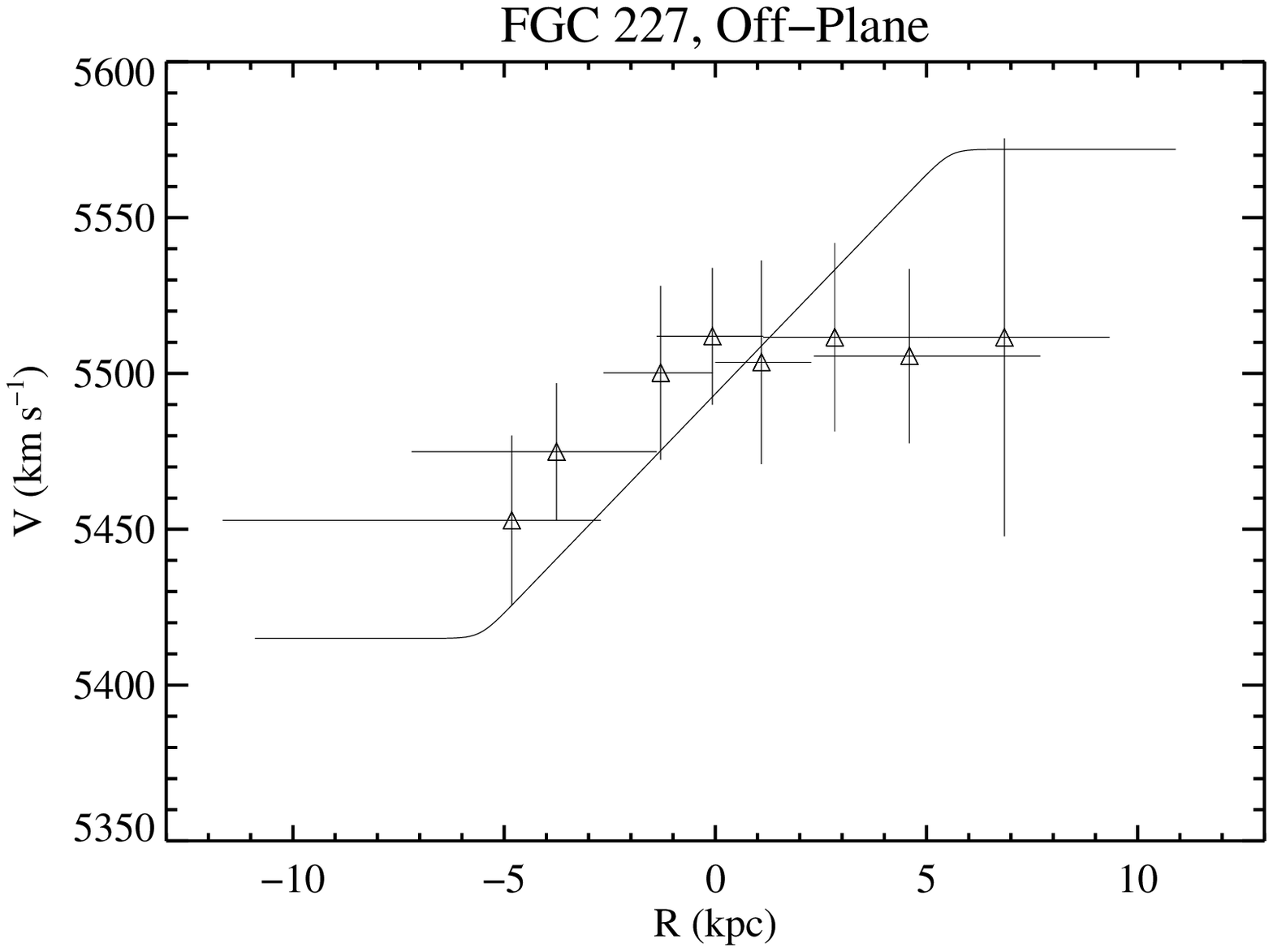}  
\caption{Extracted rotation curves for FGC~227, showing the rotation
  along the midplane (left) and above the midplane (right).  Vertical
  error bars represent the formal uncertainties of the template fitting
  while horizontal error bars represent the range of radii used in
  binning.  Points are placed at the flux-weighted mean bin position.
  The solid curve is a fit to the data points in the midplane rotation
  curve, shown for reference. Note the very slow (or complete lack of)
  rotation seen above the plane. \label{rcs_227}}
\end{figure*}

\begin{figure*}
\epsscale{1}
\plottwo{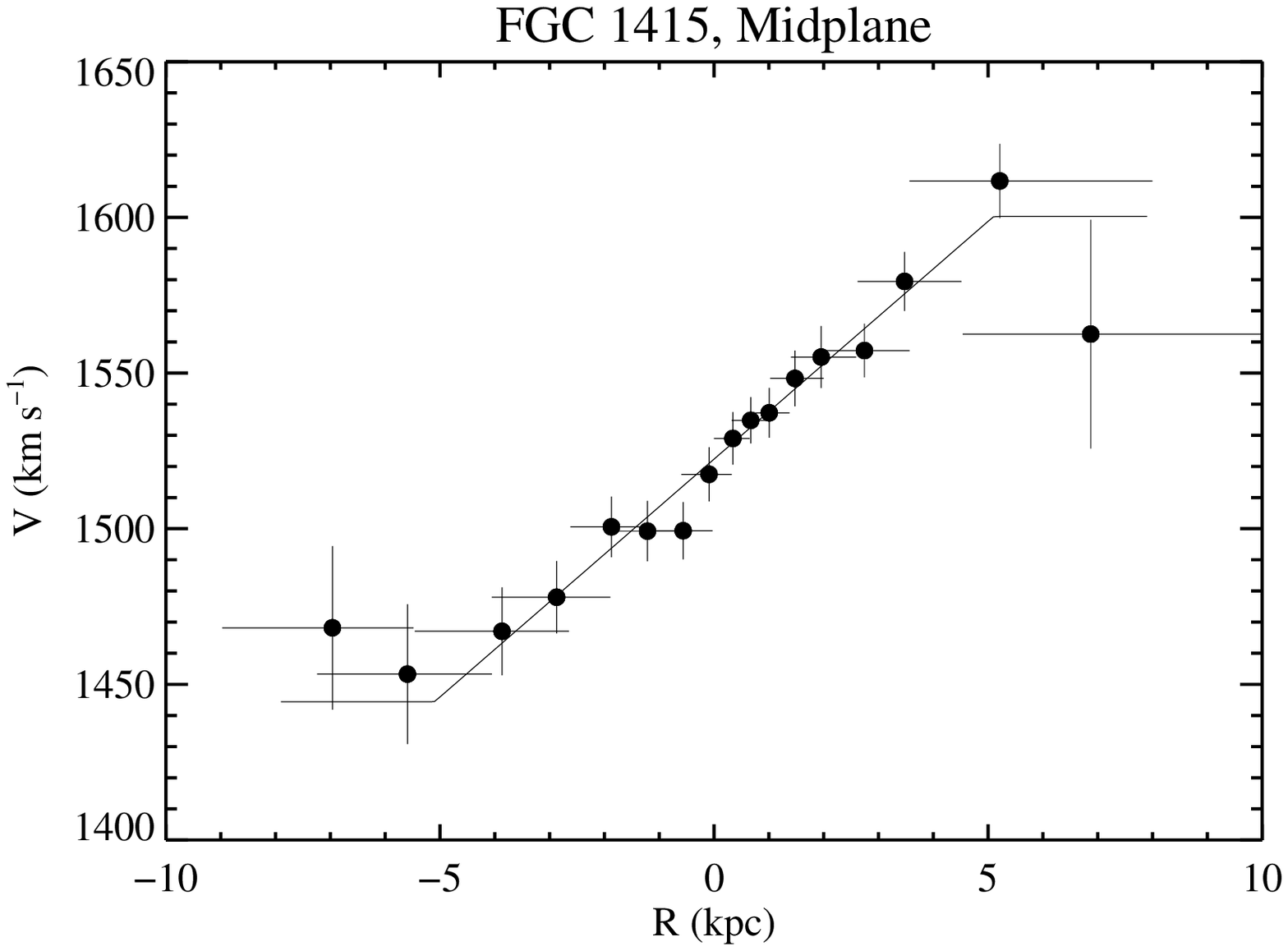}{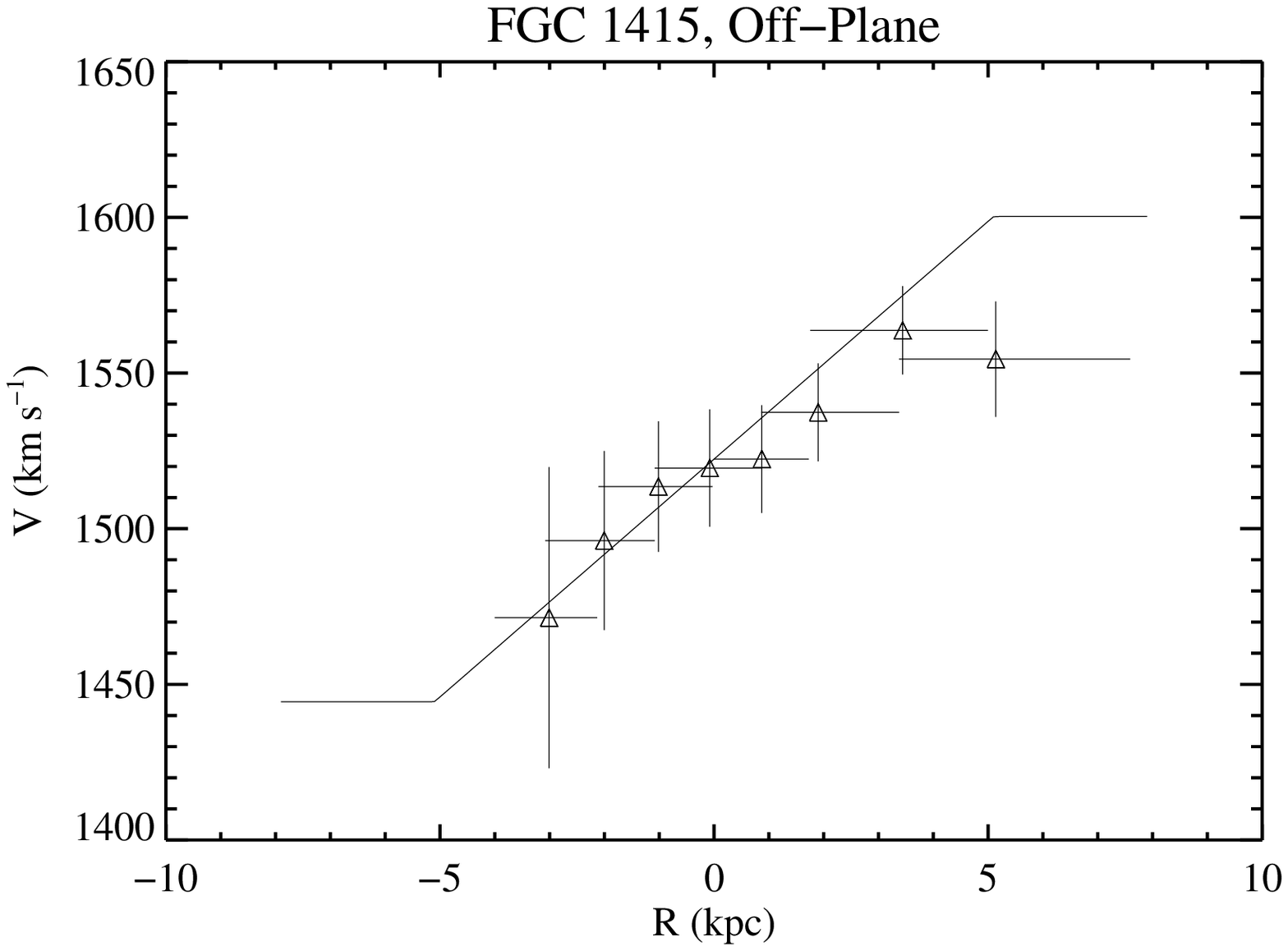}
\caption{Extracted rotation curves for FGC~1415, showing the rotation
  along the midplane (left) and above the midplane plane (right).  The
  format is the same as Figure~\ref{rcs_227}.  There are indications
  of a slight lag in the off-plane rotation curve.  \label{rcs_1415}}
\end{figure*}

The spectra of galaxies contain detailed information on their stellar
kinematics.  Methods to extract this kinematic information include
cross-correlating the galaxy spectra with a template
star~\citep{Simkin74,Tonry79, Bottema88}, or directly fitting a
redshifted stellar template~\citep{Rix92,Kelson00, Barth02}.  Direct
fitting has the advantage of permitting variable weighting across
pixels.  We therefore adopt the direct fitting method to allow us to
mask bright skylines that overlap some of the Ca {\sc ii} triplet
features.  We fit the extracted galaxy spectra with a redshifted and
broadened K-giant stellar template in the restframe region of
8480-8700 \AA, similar to the methods of \citet{Kelson00} and
\citet{Barth02}.  In particular, we construct a model galaxy spectra
$M(x)$ as
\begin{equation}
M(x)= \{ S(x+z) \otimes G(x) \} P(x)+C(x)
\end{equation}
where $S$ is the normalized template stellar spectra binned
logarithmically in wavelength, z is a redshift, $G$ is a Gaussian
broadening function, $P$ is a low-order polynomial, $C$ is a flat
continuum, and $\otimes$ denotes convolution.  The polynomial and flat
continuum are used here to correct for any errors made when
normalizing the galaxy and template spectra.  The order of the
polynomial is kept small to ensure that absorption features are not
affected.  While the broadening function can be used to measure the
line-of-sight stellar velocity dispersion, our large radial bins
introduce additional broadening into our spectra.  We therefore defer
an analysis of the stellar velocity dispersion to \S~\ref{losvdsec}, and
choose to hold the width of the Gaussian broadening function constant
during the fits.  We find the best-fit redshift and polynomial
coefficients simultaneously using a Levenberg-Marquardt routine to
minimize $\chi^2$:
\begin{equation}
\chi^2= \sum_i (O_i-M_i)\times W_i 
\end{equation}
where $O_i$ is the extracted galaxy spectra and $W_i$ is the weighting
derived from the noise spectra with masked columns set to zero weight.
Uncertainties were calculated from the covariance matrix and rescaled
so that the reduced chi-squared equals unity (i.e., we assume our fits
are of good quality).  This step is mostly conservative in that it
greatly increases error bars on several points and leaves most
virtually unchanged.  Our rescaling also prevents any residual
unmasked systematic errors from the sky subtraction from overly
influencing our error spectra and fits.  An example spectra and
template fit can be seen in Figure \ref{ex_fit}.  The raw extracted
rotation curves are presented in Figures~\ref{rcs_227}
and~\ref{rcs_1415}.

\subsection{Fitting the Rotation Curves}

Photometry suggests that our sample galaxies have both a thin and a
thick disk, both of which contribute significantly to the
measured dynamics at each slit position (Table~\ref{obstable}).  We
now attempt to decouple the kinematics of these two components.  
We first introduce a convenient analytical equation for the
rotation curves:
\begin{equation}                  \label{vceqn}
V(r)=v_0+v_c\frac{1}{(1+x^\gamma)^{1/\gamma}} \label{rc_eqn}
\end{equation}
where $v_0$ is the recessional velocity at the galaxy's center $r_0$,
$v_c$ is the asymptotic velocity at the flat part of the rotation
curve, $r_t$ is the transition radius between rising and flat sections
of the rotation curve, $\gamma$ governs the degree of sharpness in the
transition zone, and $x$ is a scaled radial parameter equal to
$r_t/(r-r_0)$.  This equation is equivalent to \citet{Courteau97}'s
Model 2 with the parameter $\beta$ set to zero.  This equation is
purely phenomenological and is flexible enough to fit a wide variety
of rotation curves.

We create model rotation curves at each slit position as follows.  We
first adopt independent rotation curves for the thick and thin disks,
$V_{thick}(r)$ and $V_{thin}(r)$, using the parameterized rotation
curve of equation~\ref{vceqn}.  We then assume our observations are
some linear combination of these two rotation curves, i.e., that each
velocity measurement is given by
\begin{equation}
V(r)=f_{thick}V_{thick}(r) + (1-f_{thick})V_{thin}(r) \label{rcw}
\end{equation}
where $f_{thick}=L_{thick}/L_{total}$ is the fraction of luminosity
from the thick disk at the slit position.  

\begin{figure*}
\epsscale{1}
\plottwo{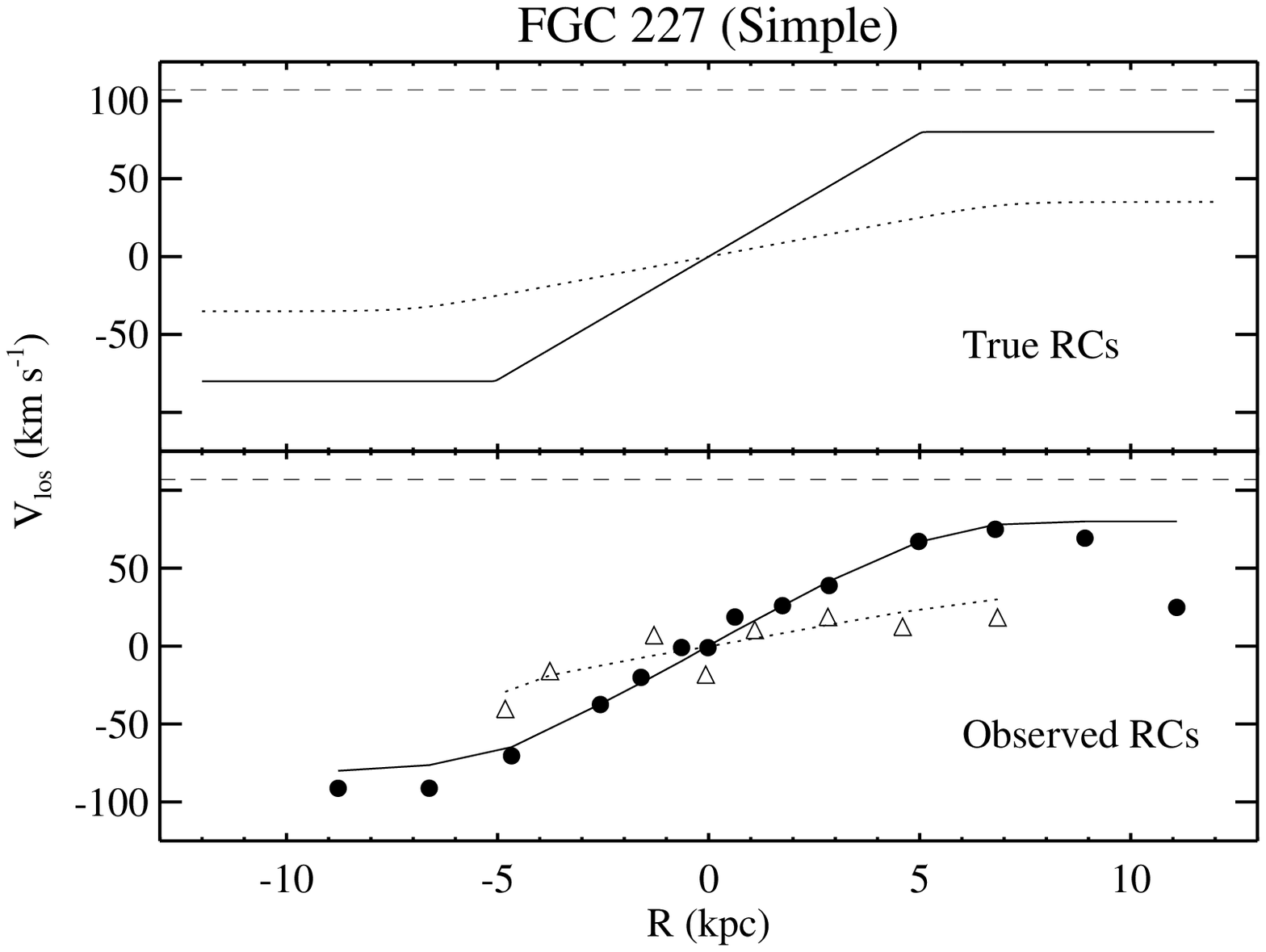}{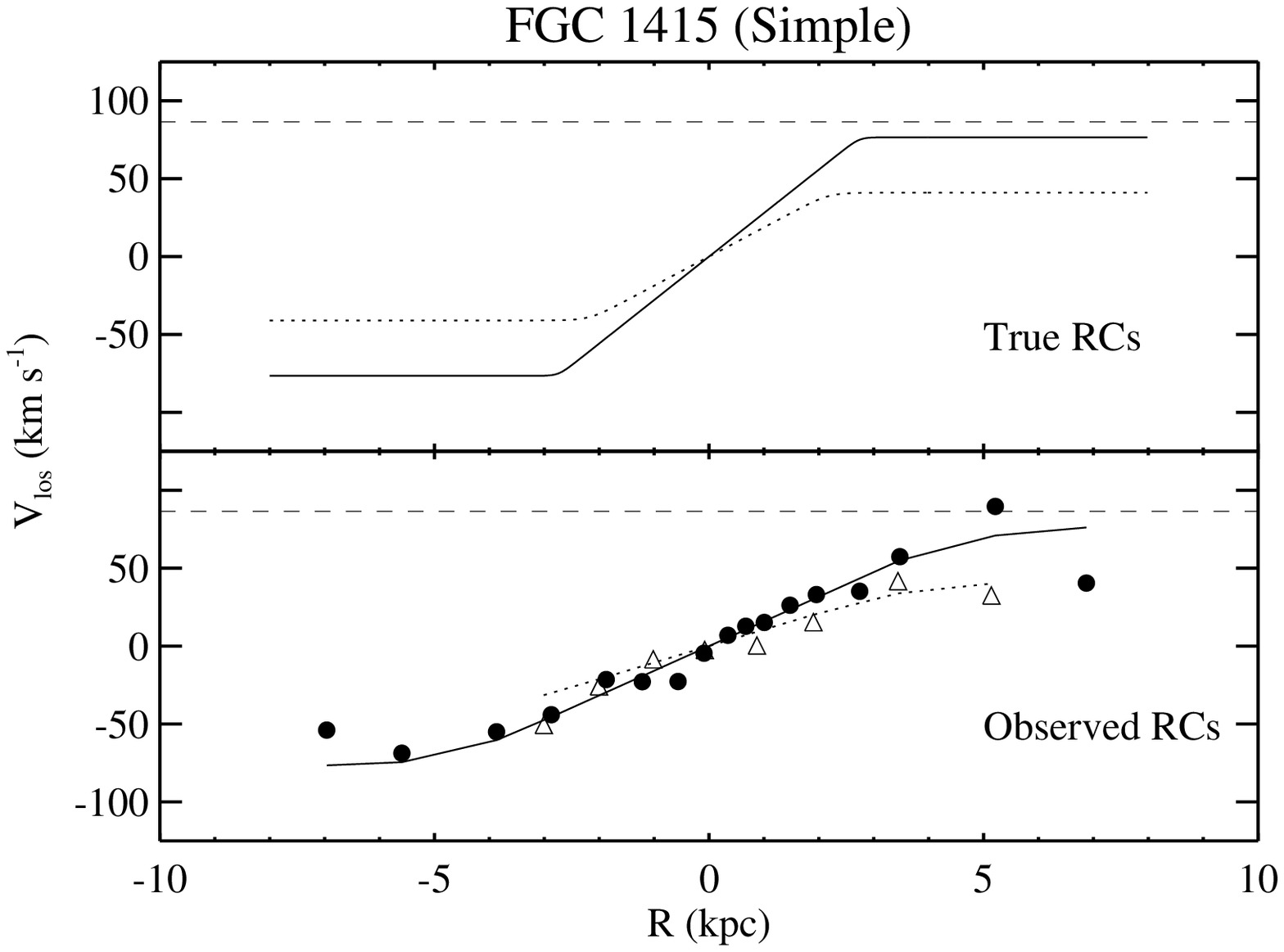} 
\caption{ Rotation curve fits to the binned rotation curves for FGC
  227 (left) and FGC~1415 (right), assuming that the midplane (solid
  line) and off-plane (dotted line) slit positions sample only the
  thin and thick disk components, respectively ($f_{thick}=0$ in the
  midplane and $f_{thick}=1$ above the plane, corresponding to the
  ``simple'' model in Tables~\ref{rc227_fits}~\&~\ref{rc1415_fits}).
  Top panels show the derived rotation curves, while bottom panels
  show those curves binned and flux-weighted identically to our
  observations.  Points in the lower panels are the same as in Figures
  \ref{rcs_227} and \ref{rcs_1415}.  Dashed lines show the W$_{50}$/2
  velocity for each galaxy. \label{simple_rcs}}
\end{figure*}

We fix the relative contributions of the thick and thin disks at each
slit position using the photometric decompositions from
\citet{Yoachim04}; as described in \S\ref{observationsec} .  We adopt
four different possible models, shown in Tables~\ref{rc227_fits} \&
\ref{rc1415_fits} and Figure~\ref{resids}.  The first is our
``preferred'' model, corresponding to our best fit photometric
decomposition.  We also include two extreme models, corresponding to
the brightest and the faintest thick disks that are compatible with
the photometric data.  For the ``faint thick'' model, we assume that
the thin disk is larger and brighter, by decreasing $\mu_{0,thin}$ and
increasing $h_{r,thin}$ and $z_{0,thin}$ by their
uncertainties, and that the thick disk is smaller and fainter, by
adjusting the same parameters in the opposite direction.  We then
recalculate $f_{thick}$.  For the ``bright thick'' model, we make the
opposite parameter shifts.  The ``faint thick'' and ``bright thick''
models are truly extreme fits to the photometric data, and result in
reduced chi-squared ($\tilde{\chi}^2$) values that are statistically
much worse than the ``preferred'' models ($\tilde{\chi}^2 -
\tilde{\chi}^2_{preferred} \gg 1$), particularly for the ``faint
thick'' model.  For completeness, we also derive a ``simple'' model
assuming that the midplane and above plane spectra are entirely
dominated by the thin and thick disks, respectively.  These latter
fits are shown in Figure \ref{simple_rcs}.  We do not include radial
gradients in $f_{thick}$, because our photometric decompositions
suggest the thick and thin disks have comparable scale lengths.
Likewise, we neglect seeing, which should have no effect on the
midplane value of $f_{thick}$.  Seeing affects the off-plane values
only marginally, shifting the value of $f_{thick}$ by $\sim\!0.05$,
which is an order of magnitude below the range of $f_{thick}$ values
we explore.

\begin{figure*}
\epsscale{1}
\plottwo{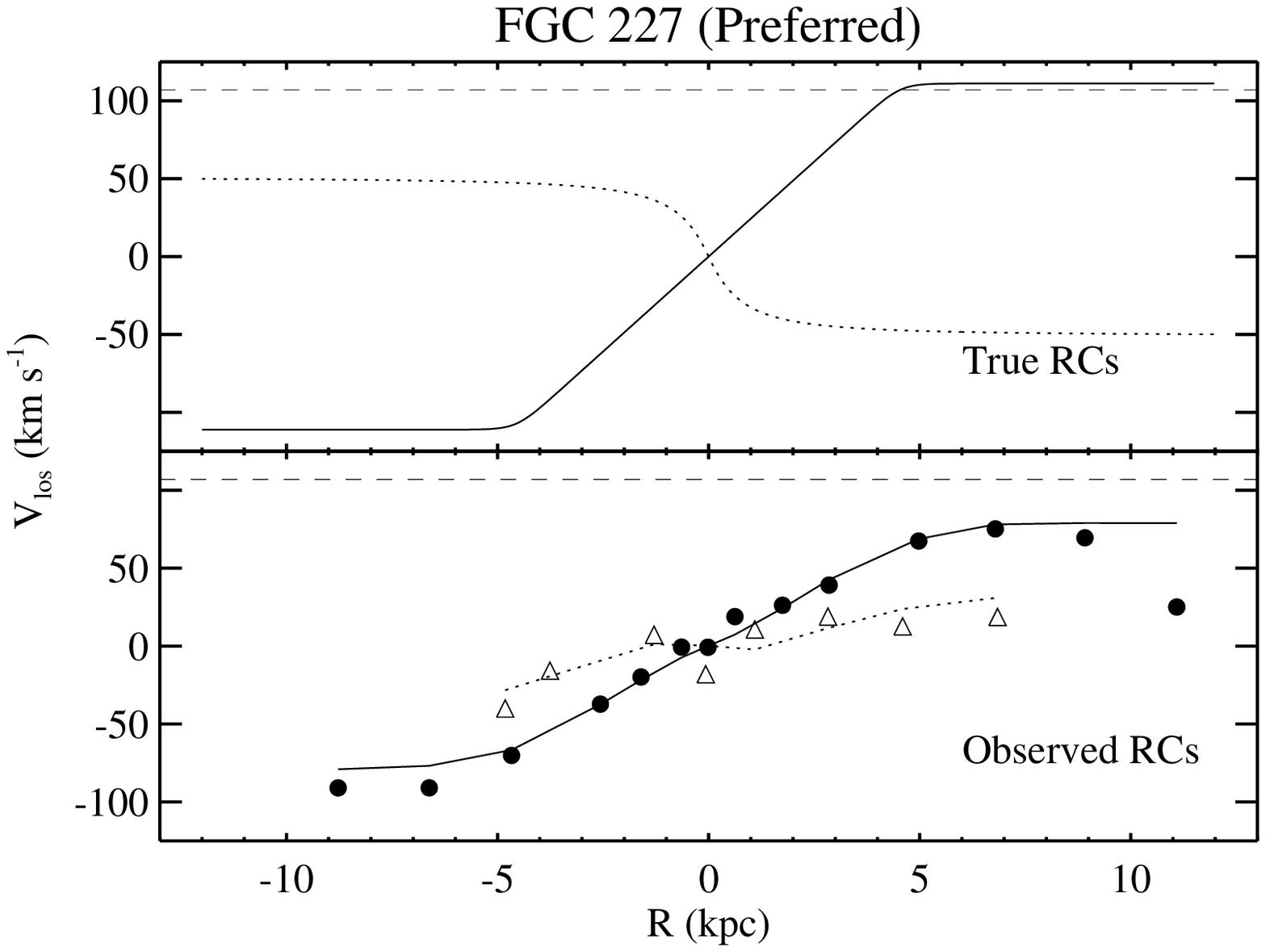}{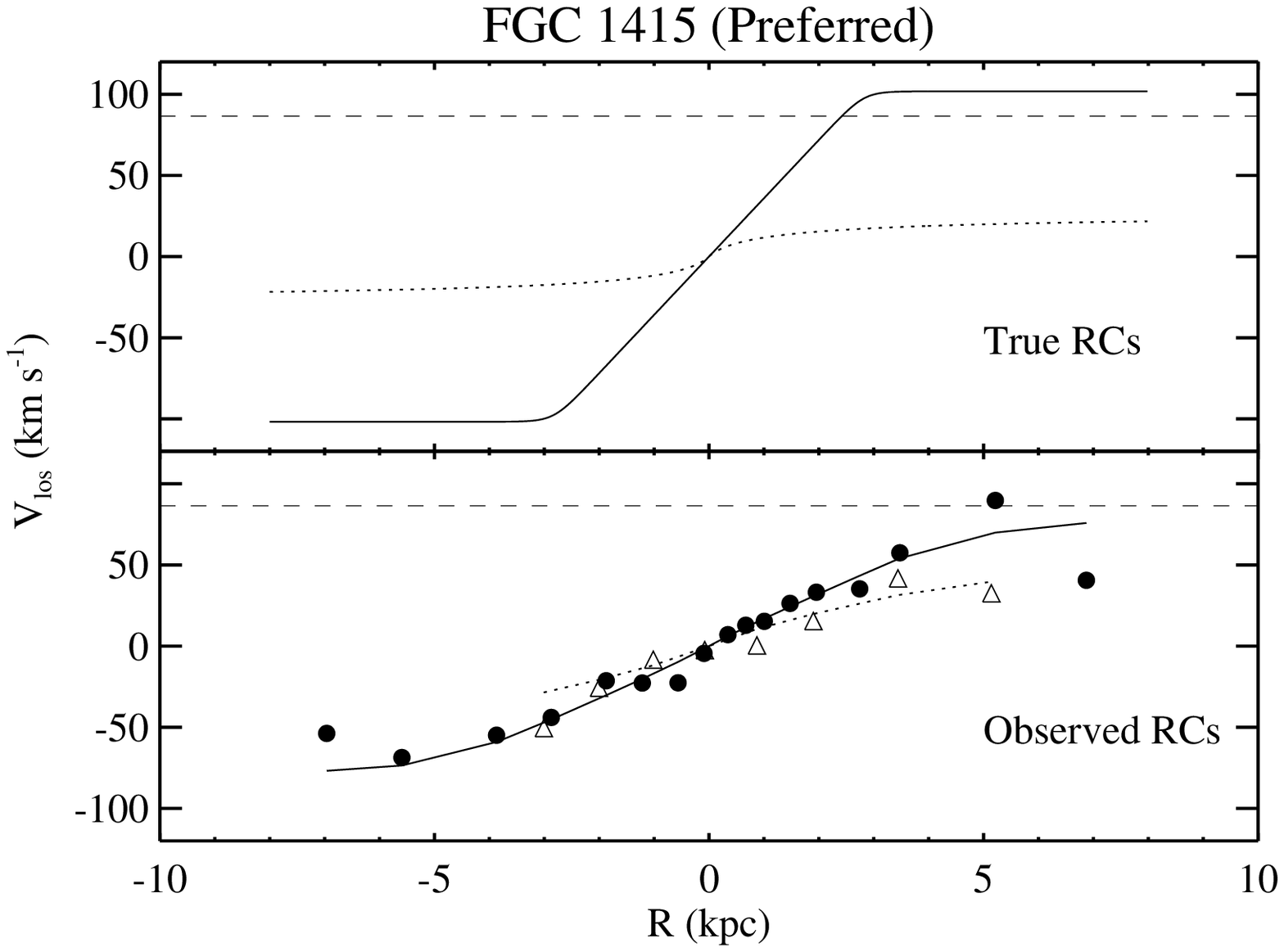}  
\caption{ Derived rotation curves using our preferred photometric
  decompositions.  For FGC~227, we assume the midplane flux is 20\%
  thick disk light while the off-plane observations are 50\% thick
  disk light.  For FGC~1415, we assume the midplane flux is 30\% thick
  disk light while the off-plane is 70\% thick disk light.  The lower
  panels display the rotation curves after they have been binned and
  flux-weighted in the same manner as our observations.  Points in the
  lower panels are the same as in Figures~\ref{rcs_227} and
  \ref{rcs_1415}. Dashed lines show the W$_{50}$/2 velocity for each
  galaxy.\label{pref_rcs}}
\end{figure*}

\begin{figure*}
\epsscale{1}
\plottwo{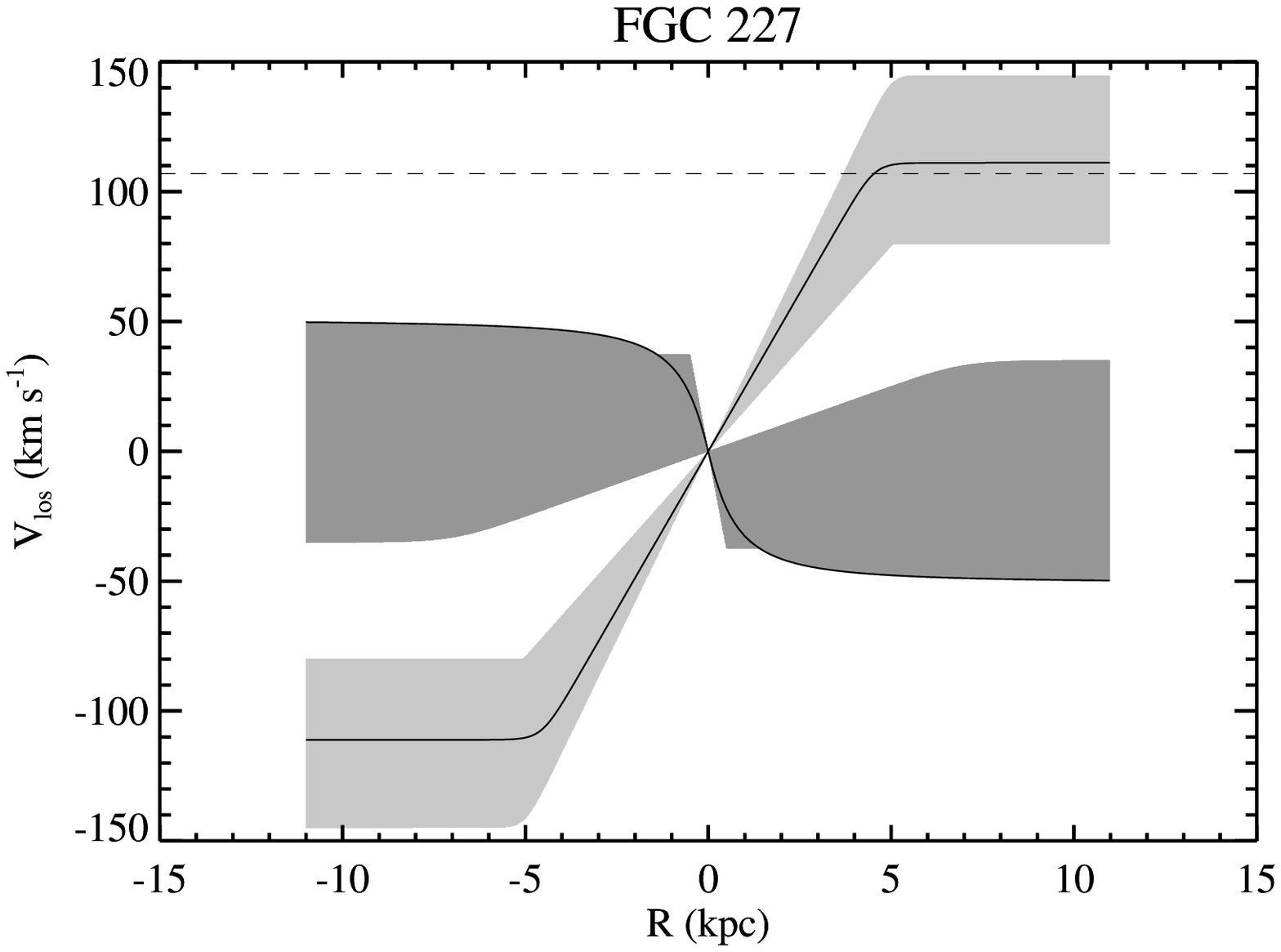}{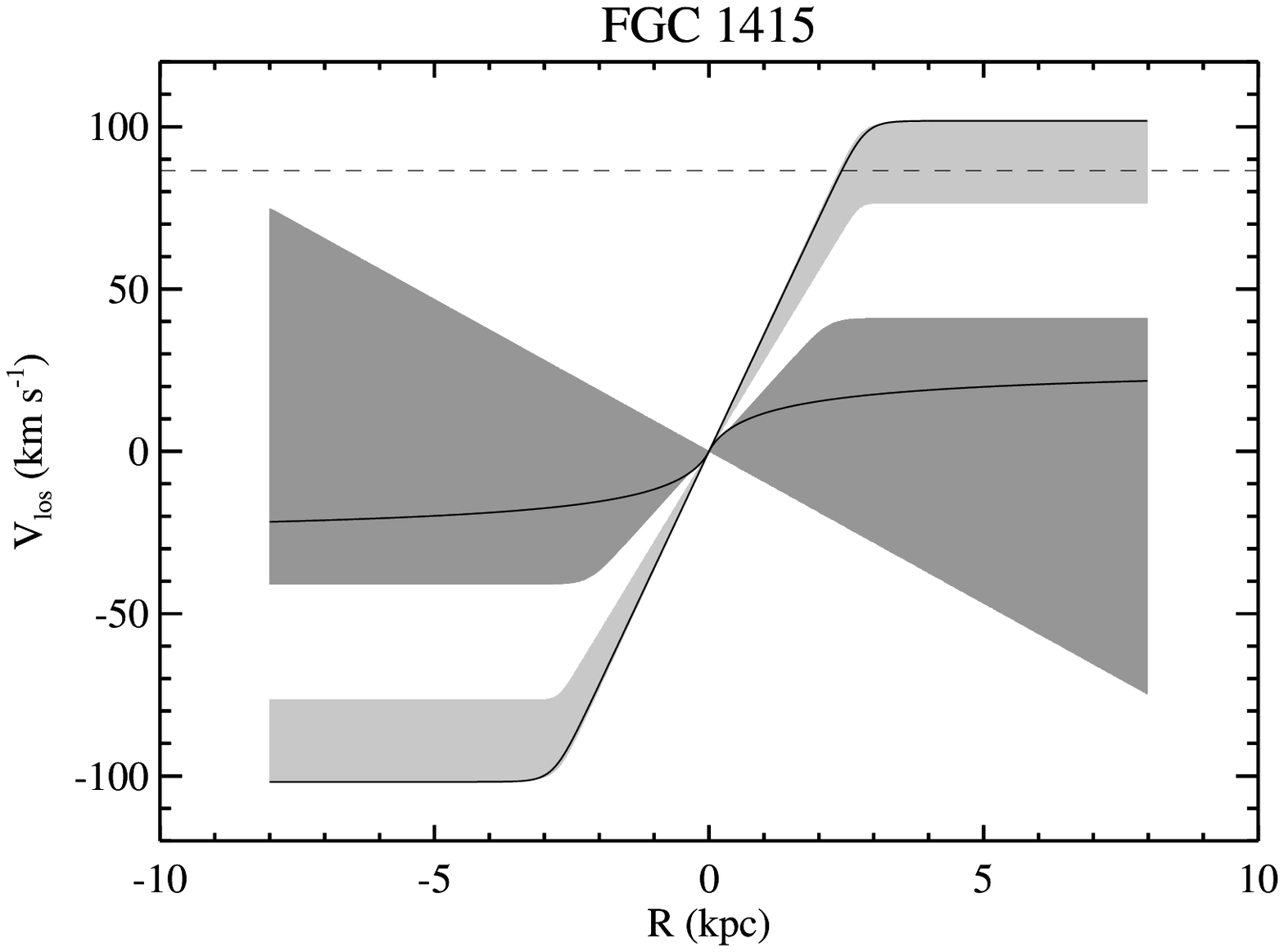}
\caption{Shaded regions showing the full range of rotation curve fits.
  The lighter region represents the full range of model thin disks,
  and the dark region shows the range of thick disks, based on
  Tables~\ref{rc227_fits} \& \ref{rc1415_fits}.  The solid lines
  represent the rotation curves calculated assuming the preferred
  $f_{thick}$ values based on photometric decomposition.
  \label{shaded_rcs}}
\end{figure*}

We bin our model rotation curves using the same bin sizes and flux
weighting as our observations.  The rotation curve parameters of both
disks are then adjusted to minimize $\chi^2$, subject to the
constraints that both rotation curves have the identical central
velocity $v_0$, that $\gamma$ is greater than 0.7, and that the
galaxy's kinematic center $r_0$ is fixed to the flux maximum of the
galaxy's continuum emission.  The value of $v_{c,thick}$ is also
constrained to be greater than -200 km s$^{-1}$ to prevent divergent
fits.  We therefore have 7 free parameters: $v_0$,
$v_{c,\mathrm{thin}}$, $r_{t, \mathrm{thin}}$,
$\gamma_{\mathrm{thin}}$, $v_{c, \mathrm{thick}}$, $r_{t,
\mathrm{thick}}$, and $\gamma_{\mathrm{thick}}$.  The best fit values
are listed in Tables~\ref{rc227_fits} \& \ref{rc1415_fits} along with
$\tilde{\chi}^2_{total}$, which is the total reduced chi-squared for
the photometric model plus rotation curve fit.  The ``faint thick''
disk models fail to converge to reasonable values, but we include the
fits for completeness.  The resulting rotation curve fits are shown in
Figures~\ref{simple_rcs}, \ref{pref_rcs} and, \ref{shaded_rcs}.  For
both galaxies, \emph{every} model has the rotation of the thick disk
lagging the thin disk, or even counter-rotating.

We now test the justification for modeling the rotation curve data
with two distinct kinematic components.  Specifically, we are
attempting to rule out the hypothesis that a single underlying
rotation curve can explain both the midplane and above plane
observations.  Using $\chi^2$ as a measure of goodness-of-fit, we
expect the minimized values of $\chi^2$ to be drawn from a chi-squared
distribution for $N-M$ degrees of freedom, where $N$ is the number of
data points and $M$ is the number of adjustable parameters in our fit.
We calculate the probability $Q$ that a value of $\chi^2$ larger than
the one we find should occur by chance.  When the midplane and
off-midplane are fit with independent rotation curves (the ``simple''
model), we find $Q$=94\% for FGC~1415 and 91\% for FGC~227, suggesting
that this is a very good fit and/or we have overestimated our
uncertainties.  We then compute the probability that the off-plane
observations are described by same simple \emph{thin} disk rotation
curve that fit the midplane, but binned in the same manner as the
thick disk observations.  This tests if both the midplane and the
off-plane observations could be sampling the same underlying rotation
curve.  We find that the off-plane observations of FGC~1415 have a
24\% chance of matching the thin disk rotation curve.  For FGC~227,
there is only a 4\% chance the off-plane observations are fit by the
simple thin disk rotation curve\footnote{These results explain the
failure of the ``faint thick'' disk models to converge.  The ``faint
thick'' disk model assumes that the thin disk dominates the light at
both slit positions, which is essentially equivalent to the hypothesis
we are ruling out.  In such a case, the thick disk is driven to
extreme values in order to minimize $\chi^2$.}.  Although these
probabilities do not formally eliminate the possibility that the
midplane and off-plane observations are due to the same underlying
rotation curve, they do suggest that it is highly unlikely.  When
coupled with the additional facts that our photometric decompositions
strongly support two disk components, that there are different colors
of the stellar populations at the two slit positions, and that the
kinematics of the thick disk and thin disk kinematics in the Milky Way
are different, we believe that our assumption of two kinematic
components is strongly justified.

A final concern is that our observations could be biased by the
effects of dust in the galaxies.  Obscuration by dust may
systematically reduce the velocities measured near the centers of
galaxies, where the optical depth due to dust is largest.  These central
velocities will be systematically lower due to the failure of
observations to penetrate to the center of the disk, allowing only
stars with small line-of-sight velocities to contribute to the
spectra.  The resulting obscured rotation curve would then appear to
be that of a rotating ring, and would thus mimic solid-body rotation
at the inner radii that are strongly obscured by dust.
\citet{Kregel04} detect just such a signature in longslit spectroscopy
of NGC 891, a massive galaxy with a prominent dust lane.

The stronger the extinction caused by dust, the more an observed
rotation curve would approach solid-body rotation \citep{Goad81}.
This could be cause for concern as our RC's are slowly rising and
fairly linear.  Using Monte Carlo radiative transfer techniques,
\citet{Matthews01} and \citet{Baes03} find that realistic optical
depths cannot explain the slowly rising rotation curves in edge-on LSB
galaxies such as ours, which lack dustlanes and have a very patchy,
highly clumped distribution of dust \citep{Dalcanton04}.
\citet{Matthews01} conclude that dust can be ignored even at
inclinations of $90\degree$ in LSB galaxies and \citet{Bosma92} find
that galaxies with rotational velocities $\sim 100$ km s$^{-1}$ are
transparent at optical wavelengths.  Our galaxies are comparable to
those in the \citet{Matthews01} and \citet{Bosma92} studies and thus
we are confident that our spectra are a true measure of stellar orbits
and not an artifact of dust extinction.

There are three additional pieces of evidence suggesting that our
results are largely unaffected by dust.  First, our galaxies are
{\emph{bluer}} in their midplanes than above them, indicating that
they do not have central dust lanes.  Second, our observations show
that the off-plane rotation curves are lagging the midplane--the
\emph{opposite} of what we would see if there was a dusty midplane.
If dust were affecting the midplane, then the thick disk would be
lagging the thin disk by even more than we have reported.  Finally, our
observations, while not in the far infrared, are red enough ($\sim8000$
\AA) that we should be able to penetrate the small amount of patchy dust
which undoubtedly exists in the galaxies even in the absence of
a concentrated dustlane.

\begin{deluxetable*}{ l c c c c c c c c c c c}
\tabletypesize{\scriptsize}
\tablewidth{0pt}
\tablecaption{FGC 227 Rotation Curve Parameters \label{rc227_fits}}
\tablehead{ 
\colhead{Model} & 
\colhead{midplane, $f_{thick}$} & \colhead{off-plane,$f_{thick}$} & 
\colhead{$v_0$} & 
\colhead{$r_{t,thin}$}  & \colhead{$r_{t,thick}$} & 
\colhead{$\gamma_{thin}$} & \colhead{$\gamma_{thick}$} & 
\colhead{$v_{c,thin}$} & \colhead{$v_{c,thick}$ } &
\colhead{$\tilde{\chi}^2_{\mathrm{total}}$}                         }
\startdata
preferred     & 0.2 & 0.5 & 5493. & 4.6 &  0.9 &  20. &   1.4 &  111. &  -51. & 5.0\\
faint thick  & 0.1 & 0.1 & 5493. & 4.8 &  0.5 & 200. & 207.5 &   87. &  -37.& 15.8 \\
bright thick & 0.5 & 0.9 & 5493. & 5.0 &  5.8 &  28. &  10.6 &  145. &   16.& 8.2 \\
simple       & 0.0 & 1.0 & 5493. & 5.1 &  7.0 & 176. &  12.0 &   80. &   35. & -- \\
\enddata
\end{deluxetable*}

\begin{deluxetable*}{ l c c c c c c c c c c c}
\tabletypesize{\scriptsize}
\tablewidth{0pt}
\tablecaption{FGC 1415 Rotation Curve Parameters \label{rc1415_fits}}

\tablehead{ 
\colhead{Model} & 
\colhead{midplane, $f_{thick}$} & \colhead{off-plane,$f_{thick}$} & 
\colhead{$v_0$} & 
\colhead{$r_{t,thin}$}  & \colhead{$r_{t,thick}$} & 
\colhead{$\gamma_{thin}$} & \colhead{$\gamma_{thick}$} & 
\colhead{$v_{c,thin}$} & \colhead{$v_{c,thick}$} &   
\colhead{$\tilde{\chi}^2_{\mathrm{total}}$}                      }
\startdata
preferred     & 0.3 & 0.7 & 1522. & 2.8 &  0.8 &  17. &  0.7 & 102. &    28.  & 24\\
faint thick  & 0.2 & 0.4 & 1522. & 2.7 & 21.4 & 100. & 15.2 &  97. &  -200. & 45  \\
bright thick & 0.4 & 0.9 & 1522. & 2.8 &  2.5 &  21. &  5.8 & 101. &    40. & 26  \\
simple       & 0.0 & 1.0 & 1522. & 2.7 &  2.2 &  33. & 13.4 &  77. &    41.  & --\\
\enddata
\end{deluxetable*}

\section{Velocity Dispersion}                               \label{losvdsec}
We now turn to an analysis of the line-of-sight velocity dispersion
(LOSVD) of our galaxy spectra.  While direct fitting returns quality
fits for recessional velocities, we find that at low S/N this method
fails when adding additional free components, and is thus unable to
measure the LOSVD.  To measure the LOSVD we have therefore used the
updated cross-correlation (XC) method of \citet{Statler95}.  This
method measures the LOSVD by fitting the peak of a template
star-galaxy XC with a broadened stellar autocorrelation function
(AXF).

To avoid possible errors from template mismatch, we use two stellar
templates, HD4388 (spectral type K3III) and HD213014 (spectral type
G9III).  Both were observed with GMOS-N for other programs and were
downloaded from the Canadian Astronomy Data Centre (CADC)
archive\footnote{Guest User, Canadian Astronomy Data Centre, which is
operated by the Herzberg Institute of Astrophysics, National Research
Council of Canada.}.  The stars were reduced using standard IRAF
routines and binned logarithmically in wavelength.  Spectra of
reference arcs were used to determine the amount of artificial
broadening needed to match the resolution of the stellar template to
the galaxy spectrum.

Before performing the XC, we performed several pre-processing steps.
First, we once again extracted galaxy spectra by summing our reduced
2D image along the spatial direction until adequate S/N was reached
($\sim10$).  Next, the galaxy and stellar templates were normalized by
dividing by a low-order fit to the continuum, after which the mean was
subtracted.  The spectra were padded (100 pixels) to eliminate
``wrap-around'' effects when cross-correlating.  Residual low
frequency components were removed by tapering the ends of the spectra in the
Fourier domain.

The faintness of our off-plane spectra requires that we bin or
observations spatially to reach adequate signal-to-noise.  However,
using such large extraction bins artificially broadens our galaxy
spectra by including stars from different parts of the rotation curve.
To eliminate this, we first run the XC fitting code to find the
recessional velocity along the galaxy.  We then shift individual rows
of the 2D galaxy spectrum to remove the measured rotation and
re-extract 1D spectra.  The XC fit is then run again in order to
measure the LOSVD.  We find no major effects from template mismatch
and can use our template stars interchangeably.  When fitting for
rotation, the two templates return nearly identical results with an
RMS spread of 1.2 km s$^{-1}$ with a maximum of 4 km s$^{-1}$
difference.  Velocity dispersions show a $\sim5$ km s$^{-1}$
systematic difference which is less than the statistical uncertainties
and may result from mismatching the dispersions based on reference arc
spectra.  

We must emphasize that the velocity dispersion measurements are very
uncertain, especially for FGC 227.  In the direct fitting method, we
were able to mask troublesome skylines and use partial sections of the
Ca triplet.  When using the XC method, masking small sections of
spectra can create large changes in the Fourier domain, so we were
forced to use only 2 of the 3 Ca absorption features for FGC 227.
Therefore, the LOSVD measurements for FGC 227 should be viewed with
caution.

\begin{figure*}
\epsscale{1}
\plottwo{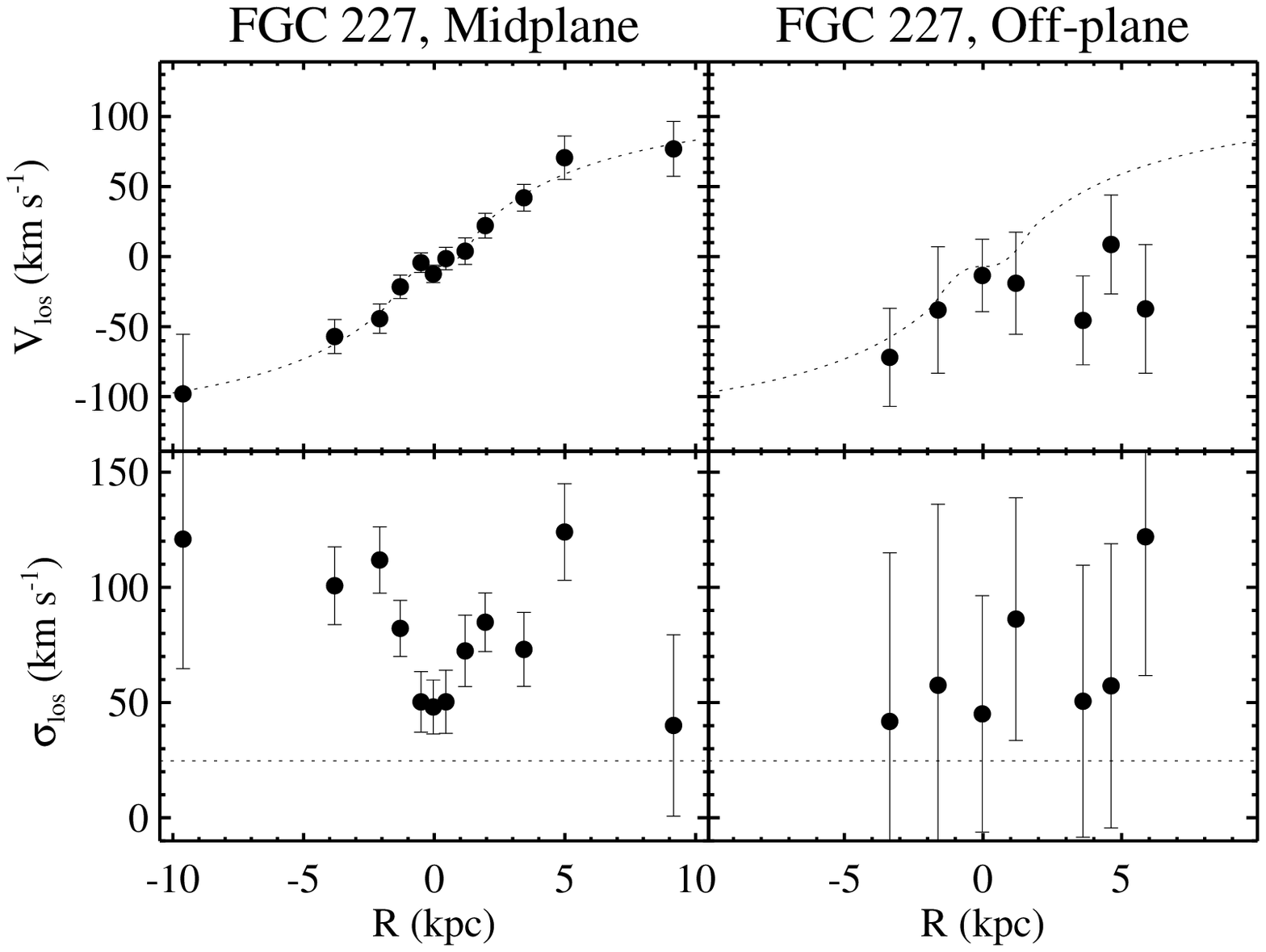}{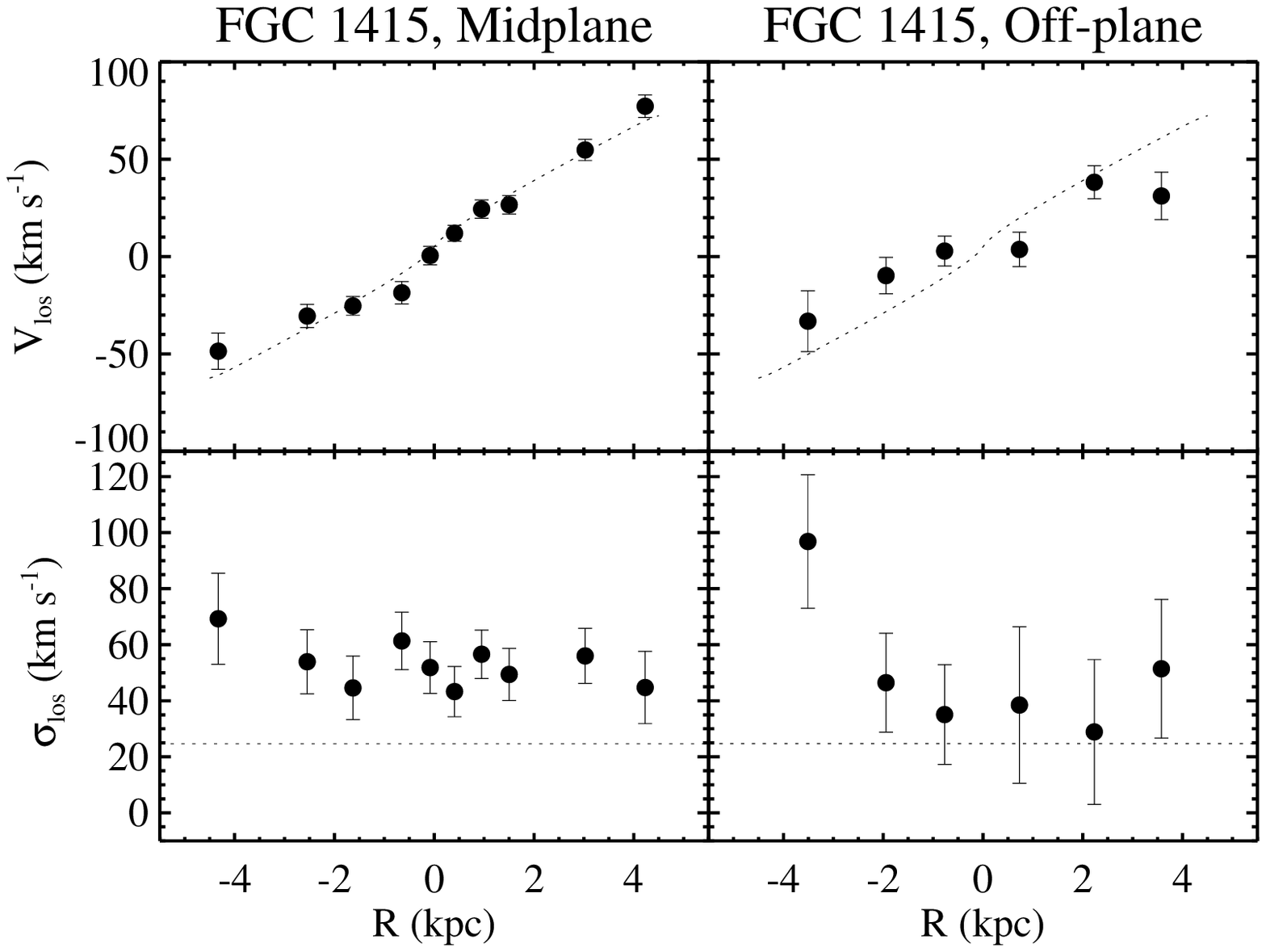}
\caption{ Results from measuring line of sight velocity and velocity
dispersions using the improved cross-correlation method.  Dashed lines
in the LOSVD plots show our pixel widths.  Dashed lines in the
velocity plots are shown for reference between midplane and off-plane
components.  Any dispersion below that level would be poorly sampled
and we would expect the XC method to fail.  The rotation measurements
are consistent with our least-squares method while FGC 227 shows a
suggestive increase in LOSVD with radius.  {\label{cc_fits}} }
\end{figure*}

The resulting measurements of the line-of-sight velocity and velocity
dispersion are shown in Figure~\ref{cc_fits}.  The rotation curves
from this method confirm the results of our direct fitting method.
Both methods show a possible slight lag for the thick disk of FGC 1415
and zero net rotation for the thick disk of FGC 227.  The LOSVD
measurements of FGC 1415 are similar for the midplane and off-plane,
and show a roughly constant velocity dispersion with radius.  The
midplane of FGC 227 shows a possible increase in the LOSVD with
radius.  Unfortunately, the off-plane observations of FGC 227 have
sufficiently low S/N and high sky-line contamination that the LOSVD
measurements have very large error bars.  We are therefore unable to
detect the increase in velocity dispersion expected for a thick disk
which is not supported by rotation.  The midplane LOSVD is
qualitatively consistent with the presence of a counter rotating thick
disk, but the off-plane observations have too low S/N to make firm
conclusions.

\begin{figure}
\plotone{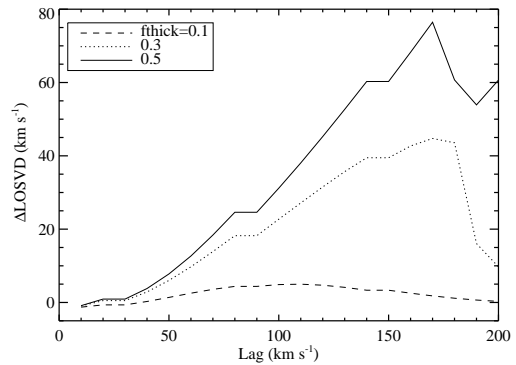} 
\caption{The expected observational effect of two disk components with
different rotational velocities on the observed LOSVD.  Plot was made
by combining two artificial spectra with intrinsic velocity
dispersions of 30 km s$^{-1}$.  For galaxies like FGC 1415 with mildly
lagging thick components (max lag $\sim80$ km s$^{-1}$ ), we would
expect a LOSVD enhancement of $\sim 15$ km s$^{-1}$ in \emph{both} the
midplane and off-plane observations.  If FGC 227 has a counter
rotating component we would expect a larger increase in LOSVD with
radius. \label{lagplot}}
\end{figure}

To understand the radial behavior of the LOSVD, we modeled the
velocity fields of two superimposed rotating disks viewed edge-on,
corresponding to a thick and thin disk rotating at two different
speeds.  We have assumed each disk has an intrinsic velocity
dispersion of $\sim30$ km s$^{-1}$.  We created a series of artificial
spectra from a combination of two broadened template stellar spectra
with different flux strengths and recessional velocities, and then
recovered their LOSVD with our XC procedure.  A sample of the
resulting LOSVD are plotted in Figure~\ref{lagplot}.  We find that, at
our resolution, the XC method is unable to separate the
cross-correlation peaks unless the two stellar components have
velocity differences $>200$ km s$^{-1}$.  For a case like FGC 1415,
where the thick disk is only mildly lagging the thin disk, we would
therefore expect only a slight increase in the observed LOSVD to occur
with radius, consistent with the constant LOSVD seen in
Figure~\ref{cc_fits}.  However, we find that adding a second component
at a different velocity can dramatically increase the observed LOSVD
if the rotational velocity difference is in the range 90-150 km
s$^{-1}$ and the second component comprises more than 20\% of the
total flux.  For FGC 227 we observe a large increase in velocity
dispersion with radius as expected if there is a counter-rotating
thick disk present.

\section{Discussion}                               \label{discussionsec}

The results of \S\ref{RCsec} and~\ref{losvdsec} indicate that the
thick disks of FGC~227 and FGC~1415 have very different kinematics.
The raw rotation curves in
Figures~\ref{rcs_227}~\&~\ref{rcs_1415} show that both galaxies appear
to be rotating more slowly above the midplane than within it.
However, when the thick and thin disk kinematics are decomposed
(Figure~\ref{shaded_rcs}), even stronger differences emerge.

In FGC~1415, the decomposition strongly favors a thick disk that
rotates in the same direction as the embedded thin disk, but with a
lag of 50\% or more.  The kinematics of FGC~1415 are therefore fully
consistent with those seen in the Milky Way, where the thick disk lags
the rotation of the thin disk by up to $100\kms$ at large scale
heights \citep{Gilmore02, Parker04}.  This suggests that any of the
theories that can explain the kinematics of the Milky Way's thick disk
can also explain the kinematics of FGC~1415.

The kinematics of FGC~227 are quite different.  The very slow rotation
above the plane in FGC~227 is due entirely to contamination from
the thin disk.  Unlike FGC~1415, the thick disk in FGC~227 is
counter-rotating, if it is even rotating at all.  While monolithic
collapse scenarios could potentially explain the rotation of FGC~1415
and the Milky Way's thick disk, they are unable to explain the slow
counter-rotation seen in FGC~227.  In such scenarios, the rotational
lag between the thick and thin disks arises from the spin up of gas as
it collapses into the thin disk while preserving its angular momentum.
Thus, the thick and thin disks must always rotate in the same
direction, in contrast to the behavior of FGC~227.  The data for
FGC~227 could marginally accommodate a very slowly co-rotating thick disk,
but only if its thick disk formed very early in the collapse process,
before the gas had spun up substantially.  This would form a thick
disk that is substantially thicker and more extended than the thin
disk.  However, the structure of FGC~227's and FGC~1415's thick and
thin disks are comparable, ruling out monolithic collapse as a
possible explanation.

The data for FGC~227 are also incompatible with models where the thick
disk forms via vertical heating of a young thin disk.  Simulations
show that infalling satellites \citep{Quinn93, Walker96, Velazquez99,
Aguerri01, Benson04} and dark matter halo substructure \citep{Font01,
Ardi03} increase the velocity dispersion of the disk stars, but do not
substantially alter their angular momenta.  The slowly rotating thick
disk of FGC~227 is therefore unlikely to have formed from a rapidly
rotating thin disk.  Another possibility is that the thick disk formed
from a thin disk that was also slowly rotating.  Such a disk must have
had a much smaller scale length because of its low angular momentum,
and thus, the heated thick disk would have a smaller scale length as
well.  This is in contrast to observations that consistently find
large thick disk scale lengths \citep{Wu02, Neeser02, Yoachim04,
Pohlen04}.  We therefore rule out vertical heating as a plausible
mechanism for forming FGC~227's thick disk.  Our conclusions are
consistent with those of \citet{Pohlen04}, who performed a very
different analysis of thick disk structural parameters in S0 galaxies.

Having ruled out monolithic collapse and vertical heating, the only
remaining possibility is that the thick disk formed from direct
accretion of stars from satellites.  In this scenario, two or more
galactic fragments merge together to form the final galaxy.  The
majority of the stars in the satellites remain in the thick disk, due
to their inability to lose kinetic energy and ``cool'' into a thinner
disk.  Meanwhile, the gas in the merging satellites collapses further
into a rotating disk, converts into stars, and forms the thin disk.
The gas in the thin disk may be supplemented at later times by
additional gas infall.  These ``shredded satellites'' are becoming
popular as explanations for a wide variety of structures in the MW
\citep{Navarro04, Martin04, Parker04, Helmi03}, as well as for low
velocity stars at high latitude in the Milky Way thick disk
\citep{Gilmore02, Soubiran03}.  This class of models can also reproduce
the enrichment patterns seen in Milky Way halo stars \citep{Chiba00,
Nissen97, Brook03, Saleh04}.

Recently, specific satellite accretion models for the formation of the
thick disk have been advocated on the basis of numerical simulations.
\citet{Abadi203} and \citet{Brook04} both identify thick disk
components in galaxies formed in cosmological simulations.  The thick
disk in \citet{Abadi203} is composed of stars accreted from merging
galactic ``sub-units'', while \citet{Brook04} find that the thick disk
stars form during the chaotic merging of gas-rich clumps that produce
the disk itself.  These new simulations form thick disks through a
series of mergers, in contrast to older models that relied on a single
final merging event to heat a thin disk into a thick disk
\citep{Quinn93, Walker96, Velazquez99}.  The merging scenario explored by the
\citet{Abadi203} and \citet{Brook04} simulations may sometimes
reproduce the behavior of the older disk heating models, for example,
when one massive satellite comes in at late times.  However, the newer
models provide mechanisms for producing a wider range of thick disk
properties.  Variations in the orbital properties and gas richness of
the merging proto-galactic clumps easily yield variations in the
kinematics of the thick disk, as well as its size and its luminosity
relative to the thin disk.

For example, satellite merging can readily produce a modest rotational
lag for the thick disk.  In general, the same satellites produce both
the gas that settles into the thin disk and the stars which are left
in the thick disk, assuming little subsequent gas accretion.  The thin
disk and thick disk should therefore show similar angular momentum
distributions, but the rotation speed of the gas should be slightly
higher, because it has contracted further into the halo.  This process
naturally leads to a thin disk which rotates more rapidly and has a
smaller scale length than the thick disk, as seen in the Milky Way and
FGC~1415.

A different combination of orbital parameters and gas richnesses in
the merging satellites can also produce the slowly rotating thick disk
of FGC~227.  If the merging satellites had little net angular momentum
(for example, two equal mass satellites orbiting the center of mass in
opposite directions), then the stars the satellites deposited in the
thick disk would show little net rotation.  However, if one of the
merging satellites was particularly gas rich, then it alone could
provide enough gas to create the rapidly rotating thin disk.  This
pathway would lead to a rapidly rotating thin disk, a stationary or
even counter-rotating thick disk, and comparable scale lengths for the
thick and thin disks, exactly as seen in FGC~227.  The resulting thick
disk would also have a large velocity dispersion, but unfortunately,
our data are not sufficient to conclusively demonstrate this for FGC
227.

One possible limitation with the stochastic satellite accretion model
is its questionable ability to consistently form highly flattened
stellar systems.  Our past imaging indicates that late-type disks
routinely host flattened thick disks with $\sim$5:1 axial ratios
\citep{Morrison97, Dalcanton02, Neeser02, Wu02}.  However, if
satellites are accreted from random directions, then there is no
{\emph{a priori}} reason why the the satellite debris should
necessarily be aligned with the embedded thin disk.  One would
therefore expect to see a significant fraction of galaxies that lack
thick disks, but are instead embedded in round stellar halos.

There are several possible resolutions to this problem.  The first is
that much of the satellite debris that is not aligned with the disk
instead becomes identified with the bulge.  This may be particularly
true when the merging satellites are gas-rich and are on largely radial
orbits, leading to centralized bursts of star formation and the formation
of a high surface brightness bulge.  Any stars formed during the burst, 
or accreted later, will tend to be identified with the brighter bulge
component.  However, this potential solution cannot explain why flattened
thick disks are seen in bulgeless late-type galaxies.

The second possibility is that the orientation of the angular momentum
vector of the accreted thick disk stars may be preferentially aligned
with that of the accreted gas that forms the thin disk, even if the
amplitude of the angular momentum vectors differ.  \citet{Abadi203}
have argued that this alignment naturally arises from the tendency of
satellites to travel along filaments, providing a preferred direction
for satellite accretion.  The alignment also may occur through
dynamical friction.  Satellites on circular orbits will have longer
lifetimes, and will thus be influenced more by dynamical friction.
Simulations by \citet{Walker96} show how a non-planar satellite orbit
can be quickly dragged into the plane by dynamical friction in only
$\sim1.5$ orbits.  The relative contribution of stars from satellites
with different orbits in the \citet{Abadi203} simulations support this
picture.

The third possibility is that the tendency
to detect flattened thick disks is a selection bias against detecting
rounder distributions of satellite debris.  If a merging satellite
does not happen to be in a coplanar orbit with the thin disk, then it
will deposit stars over a larger volume throughout the stellar halo.
The resulting distribution of stars will have a much lower surface
brightness than if the accreted stars were concentrated near the
midplane in a thick disk.  This difference naturally biases against
detecting a rounder distribution of stellar debris.  \citet{Wu02} may
have detected this component in deep intermediate-band imaging of
NGC~4565, as may have \citet{Zibetti04} using stacked images of
edge-on galaxies from the Sloan Digital Sky Survey.  However, there is
a possibility that some of this light is due to the extended light
from bulges embedded in the coadded thin disks, so this latter result
remains ambiguous.  A final possibility is that the surface brightness
of any round component will be further reduced by orbital
instabilities.  Assuming that the potential of most galaxies are
triaxial, orbits around the intermediate axes will be unstable.  These
orbits will tend to reorient around the stable short or long axes,
pulling a fraction of the stars back into a coplanar orientation.

\section{Conclusion}                    \label{conclusionsec}
 
We have measured the rotation curves and velocity dispersion of stars
at and above the midplane of two edge-on galaxies.  Comparisons
between the two positions indicate that the stars above the plane are
significantly lagging the rotation of the stars in the midplane.  We
have decomposed the observed rotation into two components,
corresponding to the thick and thin disks.  We find that the rotation
speed of the thick disk component is less than 50\% of the rotation
speed of the thin disk, and is counter-rotating in one of the two
cases. We use these observations to conclusively rule out older
monolithic collapse or disk heating models for forming all thick
disks.  We conclude that the only viable mechanism capable of forming
counter-rotating thick disks is one where the {\emph{majority}} of
thick disk stars are directly accreted from merging satellites.
Simulations suggest that the early merging required by these models is
a generic feature of hierarchical galaxy formation. As such, the
models imply that all disk galaxies should host thick disks.  This
suggests that merger models for the thick disk not only account for
the variation in kinematics, but also explains the apparent ubiquity
of thick disks around disk galaxies, as argued for by
\citet{Dalcanton02}.

We are currently expanding the sample to encompass a wider range
of galaxy masses.  We will discuss the larger sample and
the measured velocity dispersions in a future paper.

\acknowledgments

We gratefully acknowledge Karl Gebhardt for providing the K-giant
spectra that were used for cross-correlating, and discussions with
Connie Rockosi.  We'd also like to the thank the Gemini observing
staff for their skill in executing these observations.  JJD and PY
were partially supported through NSF grant CAREER AST-0238683 and the
Alfred P.\ Sloan Foundation.  Based on observations obtained at the
Gemini Observatory, which is operated by the Association of
Universities for Research in Astronomy, Inc., under a cooperative
agreement with the NSF on behalf of the Gemini partnership: the
National Science Foundation (United States), the Particle Physics and
Astronomy Research Council (United Kingdom), the National Research
Council (Canada), CONICYT (Chile), the Australian Research Council
(Australia), CNPq (Brazil), and CONICET (Argentina).



\clearpage


\end{document}